\documentclass[aps,prd,twocolumn,nomerge,superscriptaddress,preprintnumbers,superscriptaddress,nofootinbib,floatfix]{revtex4-1}
\usepackage{amsmath}
\usepackage{amssymb}
\usepackage{braket}
\usepackage[caption=false]{subfig}
\usepackage{color}
\usepackage[colorlinks=True]{hyperref}
\usepackage{longtable} 
\usepackage{graphicx}
\usepackage{textcomp}

\usepackage{pifont}% http://ctan.org/pkg/pifont

\newcommand*{\rom}[1]{\it \expandafter   \romannumeral #1}
\makeatother
\newcommand{\dcc}{LIGO-P1800391}

\graphicspath{{./images/}}

\begin{document}

\preprint{\dcc}

\title[O2CWNB]{Narrow-band search for gravitational waves from  known pulsars using the second LIGO observing run}



% LVC August 2018 author list (from LIGO-M1800242-v7)

\author{B.~P.~Abbott}
\affiliation{LIGO, California Institute of Technology, Pasadena, CA 91125, USA}
\author{R.~Abbott}
\affiliation{LIGO, California Institute of Technology, Pasadena, CA 91125, USA}
\author{T.~D.~Abbott}
\affiliation{Louisiana State University, Baton Rouge, LA 70803, USA}
\author{S.~Abraham}
\affiliation{Inter-University Centre for Astronomy and Astrophysics, Pune 411007, India}
\author{F.~Acernese}
\affiliation{Universit\`a di Salerno, Fisciano, I-84084 Salerno, Italy}
\affiliation{INFN, Sezione di Napoli, Complesso Universitario di Monte S.Angelo, I-80126 Napoli, Italy}
\author{K.~Ackley}
\affiliation{OzGrav, School of Physics \& Astronomy, Monash University, Clayton 3800, Victoria, Australia}
\author{C.~Adams}
\affiliation{LIGO Livingston Observatory, Livingston, LA 70754, USA}
\author{R.~X.~Adhikari}
\affiliation{LIGO, California Institute of Technology, Pasadena, CA 91125, USA}
\author{V.~B.~Adya}
\affiliation{Max Planck Institute for Gravitational Physics (Albert Einstein Institute), D-30167 Hannover, Germany}
\affiliation{Leibniz Universit\"at Hannover, D-30167 Hannover, Germany}
\author{C.~Affeldt}
\affiliation{Max Planck Institute for Gravitational Physics (Albert Einstein Institute), D-30167 Hannover, Germany}
\affiliation{Leibniz Universit\"at Hannover, D-30167 Hannover, Germany}
% ...

\collaboration{The LIGO Scientific Collaboration and the Virgo Collaboration}

\author{M.~Keith}
\affiliation{School of Physics and Astronomy, University of Manchester, Manchester, M13 9PL, UK}
\author{M.~Kerr}
\affiliation{Space Science Division, Naval Research Laboratory, Washington, DC 20375-5352, USA}
\author{L.~Kuiper}
\affiliation{SRON-Netherlands Institute for Space Research, Sorbonnelaan 2, NL-3584 CA Utrecht, Netherlands}
\author{A.~K.~Harding}
\affiliation{Astrophysics Science Division, NASA Goddard Space Flight Center, Greenbelt, MD 20771, USA}
\author{A.~Lyne}
\affiliation{School of Physics and Astronomy, University of Manchester, Manchester, M13 9PL, UK}
\author{J.~Palfreyman}
\affiliation{Department of Physical Sciences, University of Tasmania, Private Bag 37, Hobart, Tasmania 7001, Australia}
\author{B.~Stappers}
\affiliation{School of Physics and Astronomy, University of Manchester, Manchester, M13 9PL, UK}
\author{P.~Weltervrede}
\affiliation{School of Physics and Astronomy, University of Manchester, Manchester, M13 9PL, UK}


\date{\today}

\begin{abstract}
Isolated spinning neutron stars, asymmetric with respect to their rotation axis, are expected to be sources of continuous gravitational waves. The most sensitive searches for these sources are based on accurate matched filtering techniques, that assume the continuous wave to be phase-locked with the pulsar beamed emission. While matched filtering maximizes the search sensitivity, a significant signal-to-noise ratio loss will happen in case of a mismatch between the assumed and the true signal phase evolution. Narrow-band algorithms allow for a small mismatch in the frequency and spin-down values of the pulsar while integrating coherently the entire data set. In this paper we describe a narrow-band search using LIGO O2 data for the continuous wave emission of 33 pulsars. No evidence for a continuous wave signal has been found and upper-limits on the gravitational wave amplitude, over the analyzed frequency and spin-down ranges, have been computed for each of the targets. In this search we have surpassed the spin-down limit, namely the maximum rotational energy loss due to gravitational waves emission, for some of the pulsars already present in the O1 LIGO narrow-band search, such as J1400\textminus6325 J1813\textminus1246, J1833\textminus1034, J1952+3252, and for new targets such as J0940\textminus5428 and J1747\textminus2809. For J1400\textminus6325, J1833\textminus1034 and  J1747\textminus2809 this is the first time the spin-down limit is surpassed.
\end{abstract}

\maketitle

\section{Introduction}

Eleven gravitational wave (GW) signals have so far been detected by the LIGO \cite{0264-9381-32-7-074001,PhysRevLett.116.131103} and Virgo GW interferometers \cite{0264-9381-32-2-024001} in their first and second observing runs (O1 and O2, respectively) \cite{2018arXiv181112907T}. All the signals detected so far come from the coalescence of two compact objects. These signals belong to the class of \textit{transient signals}, since they are observed only within a short time window during the observing run.
Ten detection of binary black holes merger \cite{Abbott20162,2016PhRvL.116x1103A, 2017PhRvL.119n1101A,2017PhRvL.118v1101A,TheLIGOScientificCollaboration2017, 2018arXiv181112907T} and a detection from a binary neutron star (NS) merger \cite{2017PhRvL.119p1101A}  have been accomplished during the first and second observing runs.

Another class of GW signals potentially observable by the LIGO and Virgo detectors are the so-called \textit{continuous wave} (CW).  CWs could be potentially present during the entire data taking period of the GW detectors. Potential sources of CWs are isolated spinning NSs asymmetric with respect to their rotation axis. In the case of an oblate NS, CWs are emitted at a frequency of two times its rotational frequency.

Different types of CW searches can be performed according to the astrophysical scenario in which the NS is observed. If the NS is a pulsar, an accurate ephemeris may be available  and matched filtering techniques can be employed  to reach, ideally, the best possible sensitivity by using waveform templates that cover the entire observing run. These types of searches are referred as \textit{targeted searches}. The LIGO and Virgo Collaborations have already searched for this type of emission from  known pulsars (both isolated and some in binaries) \cite{2004PhRvD..69h2004A,2005PhRvL..94r1103A,2007PhRvD..76d2001A,2008ApJ...683L..45A,2010ApJ...713..671A,2011ApJ...737...93A,2014ApJ...785..119A,2017ApJ...839...12A, Authors:2019ztc}, for which accurate ephemerides were available.
While for NSs observed as a central compact object of a supernova remnant or in a binary system, usually accurate ephemereides are not available. 
In this case we can pinpoint the source and look for the CW signal over a wide frequency range using semi-coherent analysis, e.g. dividing the observing run in several data chunks and looking for a waveform template in each of them. Such searches are called \textit{``directed''} and offer the possibility to explore a large number of templates at the price of a lower sensitivity with respect to targeted searches\cite{2015ApJ...813...39A, 2018arXiv181211656T, 2015PhRvD..91f2008A, 2017PhRvD..95l2003A, 2017ApJ...847...47A}. Recently, there has been also a study for a possible deviation of CW signals from the General Relativity model\cite{PhysRevLett.120.031104}, by including non-tensorial modes.

Between targeted and directed searches  we find the \textit{narrow-band} searches.  Such pipelines are based on algorithms which allow to make a full coherent search and, at the same time, are able to deal with a frequency mismatch between the CW signal and the electromagnetic inferred value of the order of $500$~mHz \cite{2008ApJ...683L..45A, 2015PhRvD..91b2004A, 2017PhRvD..96l2006A}. Usually, this will correspond to the evaluation of millions of waveform templates for each pulsar considered into the analysis.

Hence, narrow-band searches offers a sensitivity comparable to the one of targeted searches while relaxing the phase-lock assumption of the CW signal with the NS rotation. The CW phase-locking is indeed a strong assumption that may prevent the detection of a CW signal. In fact, a coherent (or targeted) CW search that uses 1 year of data has a frequency resolution of about $3 \times 10^{-8}$~Hz. A mismatch between the rotational frequency inferred from the ephemeris and the CW signal frequency, of this size or larger, is enough to drastically reduce the chance of detection.

A small frequency mismatch may arise for several physical reasons, that usually are parametrized in a frequency mismatch of the form  $\Delta f_{\rm gw} \sim f_{\rm gw}(1+\delta)$ \cite{2008ApJ...683L..45A}. In the case of a differential rotation between the GW engine and the electromagnetic pulse engine, the factor $\delta$ will be proportional to the timescale of some torque which enforces correlation between the two engines. Another possibility is that the NS is freely-precessing. In this scenario the $\delta$ factor will be proportional to the angle between the star symmetry axis and the star rotation axis \cite{doi:10.1046/j.1365-8711.2002.05180.x}. In some of the previous narrow-band searches \cite{2008ApJ...683L..45A, 2015PhRvD..91b2004A} we used a value of $\delta \sim 10^{-4}$, which can accommodate the previous theoretical models. However starting from the first narrow-band search with advanced detector data \cite{2017PhRvD..96l2006A}, we explore a frequency/spin-down range corresponding to $\delta \sim 10^{-3}$. 

 Another possibility is that the pulsar ephemerides provided are not accurate enough to carry on targeted searches with the required resolution, or they are not available during the observing time of our detectors. That is the case for many low frequency and energetic pulsars observed in the X and $\gamma$-ray bands, such as J1833-1034 and J1813-1749. For these reasons, along with targeted searches, we search for CWs also with narrow-band searches.

In this paper we present the narrow-band search for CWs from 33 known pulsars using LIGO O2 data. In Sec. \ref{sec:1} we provide a brief background on the CW signal model and the algorithm used. In Sec. \ref{sec:2} we summarize the main features of the O2 narrow-band analysis, while in Sec. \ref{sec:3} we introduce the pulsars that we have selected for this search. The results of the search, followed by the upper-limits on the signal strain amplitude, are discussed in Sec. \ref{sec:4}. Finally in Sec. \ref{sec:5} we draw the conclusion of this work.

\section{Background \label{sec:1}}
\subsection{The signal}
The GW signal emitted by an asymmetric spinning NS can be written at the detector frame, using the formalism introduced in \cite{2010CQGra..27s4016A}, as the real part of 
\begin{equation}
h(t)= H_0 ( H^+ (\eta, \psi) A_+ (t) + H^\times (\eta, \psi) A_\times (t)) e^{2 \pi i f_{\mathrm{gw}} (t) t+i \phi_0}
\label{eq:Hgrande}
\end{equation}
where $f_\mathrm{gw} (t)$ is the GW frequency (which incorporates all the modulation of the signal at the detector frame) and $\phi_0$ an initial phase. The polarization amplitudes $H^+ (\eta, \psi), H^\times (\eta, \psi)$ are functions of the ratio of the polarization ellipse semi-minor to semi-major axis $\eta$ and the polarization angle $\psi$. 
The  functions $A_+ (t), A_\times (t)$ are the detector responses to the two wave polarizations. These two functions depend by the detector geographical location and the $0,\pm 1, \pm 2$ harmonics of the sidereal rotational frequency of the Earth $F_{\rm sid}$ (the inverse of the sidereal day), see \cite{2010CQGra..27s4016A} for more details. 
In Eq. \eqref{eq:Hgrande}, the amplitude of the GW $H_0$ is related to the canonical strain amplitude $h_0$ given the angle between the line of sight and the star rotation axis $\iota$:
\begin{equation}
H_0=h_0 \sqrt{\frac{1+6 \cos ^2 \iota+ \cos ^4 \iota}{4}}
\end{equation}
and
\begin{equation}
h_0=\frac{1}{d} \frac{4 \pi^2 G }{c^4} I_\mathrm{zz} f_\mathrm{gw}^2 \epsilon.
\label{eq:GW_amplitude}
\end{equation}
Being $d,~I_\mathrm{zz}$ and $\epsilon$ the star distance, moment of inertia with respect to the rotation axis and  {\it ellipticity}. The ellipticity measures the degree of asymmetry of the star with respect to its rotation axis. In the detector reference frame the signal is modulated by several effects, the most important being the \textit{R\"{o}mer delay} (also called barycentric correction) due to the detector motion, given by the Earth's orbital motion and rotation, with respect to the GW source.  Moreover the GW signal is also modulated by the source's intrinsic spin-down, due to the rotational energy loss from the source. Given a measure of the pulsar rotational frequency $f_\mathrm{rot}$, its derivative $\dot{f}_\mathrm{rot}$ and distance $d$, the GW signal amplitude can be constrained, assuming that all the star's rotational energy is lost via gravitational radiation. This theoretical value, which is an upper limit on the rotational energy that can be emitted in GWs, is called {\it spin-down limit} and is given by \cite{1998PhRvD..58f3001J}: 
\begin{equation}
h_{\rm sd}=8.06 \times 10^{-19} I_{38}^{1/2} \bigg[\frac{1\mathrm{kpc}}{d} \bigg] \bigg[\frac{\dot{f}_\mathrm{rot}}{\mathrm{Hz/s}} \bigg]^{1/2} \bigg[\frac{\mathrm{Hz}}{f_\mathrm{rot}} \bigg]^{1/2}
\label{eq:sd_limit}
\end{equation} 
where $I_{38}$ is the star's moment of inertia in units of $10^{38} \mathrm{kg \, m^2}$. Different values of the moment of inertia are possible according to the NS equation of state, mass and spin\cite{2013A&A...552A..59B}, however in this work we will assume its canonical value to be $I = 10^{38} \mathrm{kg \, m^2}$.
The corresponding spin-down limit on the star's equatorial fiducial ellipticity can be  obtained from Eq. \eqref{eq:GW_amplitude}:
\begin{equation}
\epsilon_{\rm sd}=1.91 \times 10^5 \, I_{38}^{-1/2} \bigg[\frac{\dot{f}_\mathrm{rot}}{\mathrm{Hz/s}} \bigg]^{1/2}  \bigg[\frac{\mathrm{Hz}}{f_\mathrm{rot}} \bigg]^{5/2} _.
\label{eq:eps_sd_limit}
\end{equation}
which does not depend on the star's distance.

\subsection{The 5-vector narrowband pipeline}
The narrow-band pipeline uses the 5-vector method \cite{2014PhRvD..89f2008A} and, in particular, its latest implementation for narrow-band searches described in \cite{2017CQGra..34m5007M}. 

The pipeline explores a range of frequency and spin-down values by applying barycentric and spin-down corrections to the data, and then  identifies the GW signal using its characteristic frequency components. 

The pipeline firstly removes the modulations given by the barycentric corrections and intrinsic source spindown. The barycentric corrections are applied using a frequency-independent non-uniform resampling \cite{2017CQGra..34m5007M}. The spin-down is removed by applying a phase correction on the data time series. Also the Einstein delay is corrected in the time domain. 

Once we have removed the barycentric  and spin-down modulations of a possible signal, the GW signal power is spread among five frequencies, given by the coupling of the signal frequency and the detector sidereal responses $A_+ (t), A_\times (t)$. These frequency components are: $f_{\rm gw}-2F_{\rm sid}, \,f_{\rm gw}-F_{\rm sid},\,f_{\rm gw},\,f_{\rm gw}+F_{\rm sid}$ and $f_{\rm gw}+2F_{\rm sid}$, where $F_{\rm sid}$ is the frequency corresponding to the Earth sidereal day.

Hence a pair of matched filters, one for each sidereal response function, is computed for each point of the explored parameter space. This is done using a frequency grid which allows us to compute the matched filters simultaneously over the whole analyzed frequency band. These steps are done separately for each detector. Then, the output of the matched filters, at each point of the parameter space, are combined, taking into account the phase shift\footnote{This is given by the fact that the data sampling usually does not begin at the exact same time for different detectors.} between the two data sets, in order to build a detection statistic.

The next step consists in selecting the maximum of the detection statistic for every $10^{-4}$~Hz interval and over the whole spin-down range. Within this set, points in the parameter space with a p-value below a 0.1\% threshold (taking into account the number of trials) are considered potentially interesting outliers and are subject to further analysis steps, see App.~\ref{app:thr} for more details. 

\section{The analysis \label{sec:2}}
\begin{figure*}
\centering
\includegraphics[width=1\textwidth]{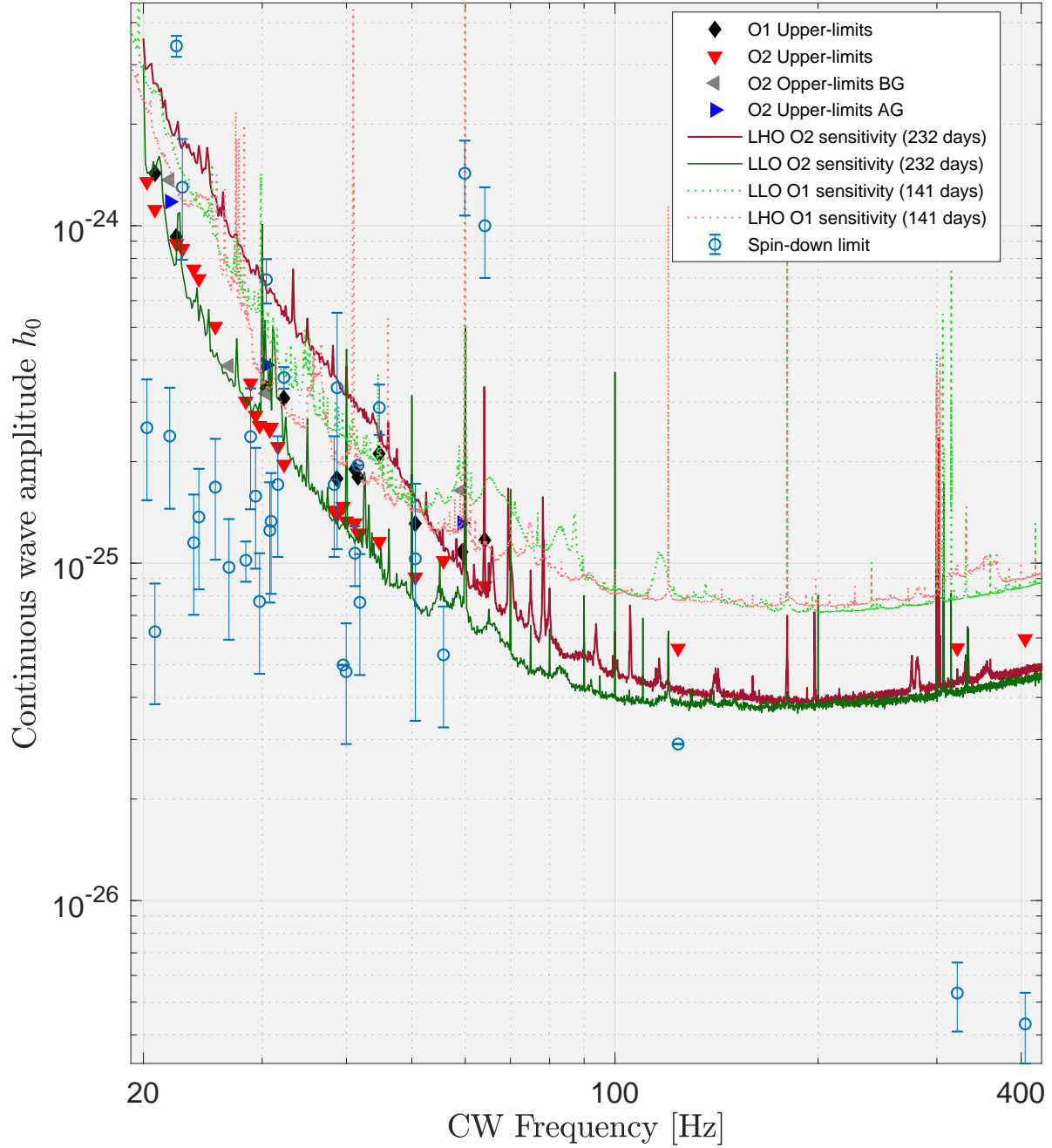}
\caption{\textit{Vertical axis:} CW amplitude, \textit{horizontal axis:} searched GW frequencies. The different lines indicate the estimated search sensitivity for O1 and O2 narrow-band searches, while the different markers indicate ULs. The labels ``AG'' and `` BG'' refers to a search performed after or before the glitch of a given pulsar. The error bars correspond to the uncertainties on the pulsar distance and correspond to $1\sigma$ confidence level.}
\label{fig:overall_plot}
\end{figure*}

The LIGO second observing run O2 started on November 30th 2016 16:00:00 UTC and ended on August 25th 2017 22:00:00 UTC, while Virgo joined the run later, on August 1st 2017 12:00:00 UTC, and ended on August 25th 2017 22:00:00 UTC.
The narrow-band search can be performed jointly between different detectors if the data sets cover the same observing time. Since Virgo O2 data covered just $\sim$1 month at the end of O2, and was characterized by a lower sensitivity with respect to LIGO data, we have decided to exclude it from the analysis.
For this analysis we have used the second version of calibratated LIGO data (C02) \cite{2017PhRvD..96j2001C}. We jointly analyzed LIGO Hanford (LHO) and LIGO Livingston (LLO) data over the period between  January 4th 2017 00:00:00 UTC and August 25th 2017 22:00:00 UTC. LLO data between the beginning of the run and December 22th 2016 have been excluded due to bad spectral contamination, while both detectors underwent a commissioning break between December 22th 2016 and January 4th 2017. The observing time $T_{\rm obs}$ was $\sim232$ days, implying frequency and spin-down bins of, respectively, $\delta f=5 \times 10^{-8}$~Hz and  $\delta \dot{f}=2.5 \times 10^{-15}$~Hz/s.
LHO and LLO duty cycles were about 45\% and the  56\% and corresponded to an effective observing time of 104 days and 129 days respectively\footnote{With the exception of pulsars that have glitched during the analysis. For those we have performed two independent analysis before and after the glitch.}.
The sensitivity of the O2 search is reported in Fig. \ref{fig:overall_plot}, where we show also O1 sensitivity. 
While at lower frequency only O2 LLO seems to be much better than O1, at higher frequencies the sensitivity is significantly better for both the detectors. In order to validate the analysis, we have looked for 4 hardware injections in the data checking if their  parameters were recovered correctly, see Appendix A.

The explored frequency and spindown ranges were set to 0.4\% of the pulsar rotational frequency and spindown reported in the ephemeris. 
Since in this analysis we sub-sampled data at $1$~Hz, the explored frequency region of some pulsars has been chosen manually in order to avoid a possible signal aliasing.

We have decided to select as {\it outliers} for the follow-up the points in the parameter space with a value of the detection statistic corresponding to a p-value of $0.1\%$ (taking into account the number of trials) or smaller. In the previous O1 search  we used a threshold of $1\%$, due to the fact that data quality of LHO and LLO was significantly different at lower frequencies, see Appendix B for more details.

\section{Selected targets \label{sec:3}}

In our O2 analysis we have selected as an initial set of targets all the pulsars present in the O1 narrow-band search\cite{2017PhRvD..96l2006A}. Then we have enlarged it, deciding to analyze all the pulsars with rotation frequency of 10~Hz and 350~Hz with spin-down limit, given in Eq. \eqref{eq:sd_limit}, within a factor 10 from the optimal sensitivity of the search of O2 LLO (in most cases).
This choice has been driven by the fact that available pulsar  distances can be affected by a large error. The spindown limit has been computed according to the most recent estimation of the distance given in the ATNF catalog\cite{2005AJ....129.1993M} (\textit{v1.58}) and extrapolating the rotational frequency and spindown rate at  the O2 epoch.  For the pulsars:  J1028\textminus5819, J1112\textminus6103, J1813\textminus1246 and J2043+2740, we have checked that the extrapolated rotational parameters together with the ranges explored in the narrow-band search covers the values reported by the updated ephemeris during the O2 epoch in \cite{Authors:2019ztc}. While for the pulsars J0835\textminus4510, J0940\textminus5428, J1105\textminus6107, J1410\textminus6132, J1420\textminus6048, J1531\textminus5610, J1718\textminus3825, J1809\textminus1917 and J1838\textminus0655 the extrapolated spin-down rate resulted off-range with respect to the one reported in \cite{Authors:2019ztc} and for this reason searched parameter space has been  adjusted in such a way to cover the updated values.
For the pulsars J0205+6449, J0534+2200, J1913+1011, J1952+3252 and J2229+6114 we have used updated ephemerides provided by the telescopes at Jodrell Bank (UK).
For the remaining pulsars, no monitoring is present during the O2 run. Even though we are aware that an extrapolation from outdated ephemerides might bring to a GW search which does not  cover the actual pulsar rotational parameters during O2, we have decided to carry on the analysis in such a way to exploit the possibility that the actual pulsar rotational parameters were covered even partially by the narrow-band search.

 Tab. \ref{tab:dist_unc} reports the spindown limit on amplitude $h_0$ and ellipticity $\epsilon$ for each target, given their distance estimation and uncertainty. Hereafter, the distance uncertainties are propagated to the derived quantities (such as the spin-down limit) assuming normal distributions, namely:
\begin{equation*}
\sigma^2_{Y}=\bigg(\frac{\partial Y}{\partial d}\bigg)^2 \sigma^2_d,
\end{equation*}
with $Y$ being a function of the distance and $\sigma^2$ the distribution variance.
\begin{table}
\caption{\label{tab:dist_unc} Properties of analyzed pulsars. The second column reports the distance as provided by the ephemerides, based on the dispersion measure and the Galactic electron density model of \cite{2017ApJ...835...29Y}. If the pulsar distance is estimated according to an independent measure, this is referred next to the name entry. The distance uncertainty refers to $1\sigma$ confidence level and is assumed to have a normal distribution. In the third and fourth column the spin-down limit $h_{\rm sd}$ and the corresponding ellipticity $\epsilon_{\rm sd}$ are computed using Eqs. \eqref{eq:sd_limit}-\eqref{eq:eps_sd_limit}.}
\begin{tabular}{l|ccc}
\textbf{Name}& $d$ [kpc]&$h_{\rm sd}$& $\epsilon_{\rm sd}$ \\ \hline
J0205+6449\cite{2013AandA...560A..18K}& $2.0 \pm 0.3$ &$(6.9 \pm 1.1) \cdot 10^{-25}$ & $1.42 \cdot 10^{-3}$ \\
J0534+2200\cite{2008ApJ...677.1201K}& $2.0 \pm 0.5$ &$(1.4 \pm 0.4) \cdot 10^{-24}$ & $7.56 \cdot 10^{-4}$ \\
J0537\textminus6910\cite{2013Natur.495...76P}& $49.7 \pm 0.2$ &$(2.91 \pm 0.02) \cdot 10^{-26}$ & $8.90 \cdot 10^{-5}$ \\
J0540\textminus6919\cite{2013Natur.495...76P}& $49.7 \pm 0.2$ &$(4.99 \pm 0.02) \cdot 10^{-26}$ & $1.50 \cdot 10^{-3}$ \\
J0835\textminus4510\cite{2003ApJ...596.1137D}& $0.28 \pm 0.02$ &$(3.4 \pm 0.3) \cdot 10^{-24}$ & $1.80 \cdot 10^{-3}$ \\
J0940\textminus5428& $0.4 \pm 0.2$ &$(1.3 \pm 0.5) \cdot 10^{-24}$ & $8.97 \cdot 10^{-4}$ \\
J1028\textminus5819& $1.4 \pm 0.6$ &$(2.4 \pm 1.0) \cdot 10^{-25}$ & $6.70 \cdot 10^{-4}$ \\
J1105\textminus6107& $2.4 \pm 0.9$ &$(1.7 \pm 0.7) \cdot 10^{-25}$ & $3.82 \cdot 10^{-4}$ \\
J1112\textminus6103& $4.5 \pm 1.8$ &$(1.3 \pm 0.5) \cdot 10^{-25}$ & $5.61 \cdot 10^{-4}$ \\
J1300+1240\cite{2012ApJ...755...39V}& $0.7 \pm 0.2$ &$(5.3 \pm 1.3) \cdot 10^{-27}$ & $3.17 \cdot 10^{-8}$ \\
J1302\textminus6350& $2.3 \pm 0.9$ &$(7.6 \pm 3.0) \cdot 10^{-26}$ & $9.52 \cdot 10^{-5}$ \\
J1400\textminus6325\cite{2010ApJ...716..663R}& $0.9 \pm 0.3$ &$(1.0 \pm 0.3) \cdot 10^{-24}$ & $2.07 \cdot 10^{-4}$ \\
J1410\textminus6132& $13.5 \pm 5.3$ &$(4.8 \pm 1.9) \cdot 10^{-26}$ & $3.83 \cdot 10^{-4}$ \\
J1420\textminus6048& $5.6 \pm 2.2$ &$(1.6 \pm 0.7) \cdot 10^{-25}$ & $9.81 \cdot 10^{-4}$ \\
J1524\textminus5625& $3.4 \pm 1.3$ &$(1.7 \pm 0.7) \cdot 10^{-25}$ & $8.25 \cdot 10^{-4}$ \\
J1531\textminus5610& $2.8 \pm 1.1$ &$(1.2 \pm 0.5) \cdot 10^{-25}$ & $5.47 \cdot 10^{-4}$ \\
J1617\textminus5055& $4.7 \pm 1.9$ &$(2.4 \pm 1.0) \cdot 10^{-25}$ & $1.28 \cdot 10^{-3}$ \\
J1718\textminus3825& $3.5 \pm 1.4$ &$(9.7 \pm 3.8) \cdot 10^{-26}$ & $4.48 \cdot 10^{-4}$ \\
J1747\textminus2809& $8.2 \pm 3.2$ &$(1.7 \pm 0.7) \cdot 10^{-25}$ & $8.97 \cdot 10^{-4}$ \\
J1747\textminus2958& $2.5 \pm 1.0$ &$(2.5 \pm 1.0) \cdot 10^{-25}$ & $1.47 \cdot 10^{-3}$ \\
J1809\textminus1917& $3.3 \pm 1.3$ &$(1.4 \pm 0.6) \cdot 10^{-25}$ & $7.27 \cdot 10^{-4}$ \\
J1811\textminus1925& $5.0 \pm 2.0$ &$(1.3 \pm 0.6) \cdot 10^{-25}$ & $6.59 \cdot 10^{-4}$ \\
J1813\textminus1246\cite{2014ApJ...795..168M}& $>2.5$ &$<1.9 \cdot 10^{-25}$ & $2.67 \cdot 10^{-4}$ \\
J1813\textminus1749\cite{2012ApJ...753L..14H}& $4.7 \pm 0.8$ &$(2.9 \pm 0.5) \cdot 10^{-25}$ & $6.42 \cdot 10^{-4}$ \\
J1831\textminus0952& $3.7 \pm 1.5$ &$(7.7 \pm 3.0) \cdot 10^{-26}$ & $3.04 \cdot 10^{-4}$ \\
J1833\textminus1034\cite{2006ApJ...637..456C}& $4.1 \pm 0.3$ &$(3.6 \pm 0.3) \cdot 10^{-25}$ & $1.32 \cdot 10^{-3}$ \\
J1838\textminus0655\cite{2008ApJ...681..515G}& $6.6 \pm 0.9$ &$(1.0 \pm 0.2) \cdot 10^{-25}$ & $7.94 \cdot 10^{-4}$ \\
J1913+1011& $4.6 \pm 1.8$ &$(5.4 \pm 2.1) \cdot 10^{-26}$ & $7.54 \cdot 10^{-5}$ \\
J1952+3252\cite{2012ApJ...755...39V}& $3.0 \pm 2.0$ &$(1.0 \pm 0.7) \cdot 10^{-25}$ & $1.15 \cdot 10^{-4}$ \\
J2022+3842\cite{2014ApJ...790..103A}& $10.0 \pm 2.0$ &$(1.1 \pm 0.3) \cdot 10^{-25}$ & $6.00 \cdot 10^{-4}$ \\
J2043+2740 & $1.5 \pm 0.6$ &$(6.3 \pm 2.5) \cdot 10^{-26}$ & $2.03 \cdot 10^{-4}$ \\
J2124\textminus3358\cite{2016MNRAS.458.3341D}& $0.4 \pm 0.1$ &$(4.3 \pm 1.0) \cdot 10^{-27}$ & $9.49 \cdot 10^{-9}$ \\
J2229+6114\cite{2001ApJ...552L.125H}& $3.0 \pm 2.0$ &$(3.3 \pm 2.3) \cdot 10^{-25}$ & $6.27 \cdot 10^{-4}$ \\

\end{tabular}
\end{table}

The spindown limits are compared to the estimated narrow-band search sensitivity  in Fig \ref{fig:overall_plot}. The analysis covers the 11 targets that we have  already analyzed for O1 plus 22 new targets. Based on the estimated sensitivity we expected to surpass the spin-down limit, in the O2 analysis, for 9 of the 11 O1 targets. The exceptions are J2043+2740 and J2229+6114, for which the current distance estimation has been increased with respect to the ATNF catalog \textit{v1.54} (the catalog used for O1 \cite{2017PhRvD..96l2006A}).

The new O2 targets mainly consist of pulsars with rotational frequencies within $10$~Hz and $20$~Hz with spin-down rate $<-10^{-12}$~Hz/s, but there are also a few  millisecond pulsars, for which we can approach the spin-down limit. Among these there is the millisecond pulsar J2124+3358, for which we expect to barely approach the spin-down limit with targeted searches. One of these millisecond pulsars, J1300+1240, is located in a binary system. However, according to the orbital parameters in the ephemeris, the intrinsic binary orbital modulation on a possible CW signal would be of the order of $\Delta f_{\rm bin}\approx 10^{-10}$~Hz, that is below our frequency resolution and hence can be neglected\footnote{The frequency shift due to the binary motion has been computed using \cite{2015PhRvD..91j2003L}.}.
Millisecond pulsars are characterized by a low rotational spindown value $\dot{f}_{\rm rot}$ together with a high rotational frequency $f_{\rm rot}$, hence according to Eq. \eqref{eq:sd_limit}, also their spindown limit will be harder to surpass by our search sensitivities.
Although the narrow-band search is currently not  sensitive enough for the millisecond pulsars, we have decided to perform the search in order to test the capabilities of the pipeline at higher frequencies.

Furthermore, pulsars J0205+6449, J0534+2200, J0835-4510,J1028\textminus5819 and J1718\textminus3825 had a glitch during O2. J0205+6449 glitched on May 27th 2017, J0534+2200 glitched on Mar 27th 2017, J0835-4510 had a glitch on Dec 16th 2016 \cite{2018Natur.556..219P}, J1028\textminus5819 glitched on May 29th 2017 and J1718\textminus3825 glitched on May 1st July 2017 \cite{Authors:2019ztc}. For these pulsars we have performed two independent analyses, one before and one after the glitch, excluding the day in which the glitch was present.
For J0835-4510 and J1718\textminus3825 only the analysis after or before the glitch has been done, since few data were available before or after the two glitches.

Tab. \ref{tab:O2space} reports the frequency/spin-down regions that we have analyzed for each of the 33 targets. The reference time for the rotational parameters of the pulsars is  December 1 2016 00:00:00 UTC. 

%Concerning the pulsar J1838\textminus0655, at the time  this narrow-band analysis was carried out, we used as ephemerides the one provided  by the ATNF catalog \textit{v1.58} with epoch MJD 54522 (2008) and extrapolated the ephemeris to the reference time used in this analysis (December 2016). Recently we received  new ephemeris by Swift and NICER\footnote{L. Kuiper and A. K. Harding private communication.} covering a period  between 2017 Mar 17th to 2018 Oct 13th  with epoch of 2018 June 13th.  The GW frequency inferred by this new ephemeris is well covered by the narrow-band search performed. On the other hand the search setup from from the ATNF ephemeris does not cover the $\dot{f}$ value inferred from the recent NICER ephemeris. In fact, the closest  spin-down  values in the  two possible narrow-band searches are $\sim 15$ spin-down bins apart ($\approx 3.75 \cdot 10^{-14}$ Hz/s).
%However, since we do not have an accurate model of the possible mismatch between the GW and  electromagnetic pulse inferred spin-down, and since the two spin-down spaces are very close to each other, we have decided to show the analysis of this pulsar for completeness.

\begin{table*}[h]
\centering
\caption{\label{tab:O2space} First column: pulsar name. Second and third columns: central frequency and  frequency width  explored in the search. Fourth and fifth columns: central spin-down and spin-down ranges explored in the search. Sixth and seventh column: number of templates in frequency and spin-down. Frequency and spin-down resolutions are, respectively, $\delta f \sim 5 \times 10^{-8}$~Hz, $\delta\dot{f}  \sim 2.5 \times 10^{-15}$~Hz/s. The labels ``AG'' and ``BG'' indicate, respectively, after and before the glitch. Note that the frequency and spin-down resolution, and hence the number of templates, is lower in the case of pulsars with a glitch.}
\begin{tabular}{l|cccccc}
\textbf{Name}& $f$ [Hz]&$\Delta f$ [Hz]& $\dot{f}$ [Hz/s]&$\Delta \dot{f}$ [Hz/s]&$n_{f}$[$10^6$]&$n_{\dot{f}}$ \\ \hline
J0205+6449 AG & 30.41 & 0.06 & $-8.61 \cdot 10^{-11}$ & $2.72 \cdot 10^{-13}$ & $0.47$ & 17 \\ 
J0205+6449 BG & 30.41 & 0.06 & $-8.61 \cdot 10^{-11}$ & $2.44 \cdot 10^{-13}$ & $0.74$ & 37 \\  
J0534+2200 AG & 59.30 & 0.12 & $-7.38 \cdot 10^{-10}$ & $1.50 \cdot 10^{-12}$ & $1.53$ & 251 \\ 
J0534+2200 BG & 59.30 & 0.12 & $-7.38 \cdot 10^{-10}$ & $1.56 \cdot 10^{-12}$ & $0.82$ & 75 \\ 
J0537\textminus6910 & 123.86 & 0.25 & $-3.92 \cdot 10^{-10}$ & $8.01 \cdot 10^{-13}$ & $4.95$ & 321 \\ 
J0540\textminus6919 & 39.39 & 0.08 & $-3.71 \cdot 10^{-10}$ & $7.56 \cdot 10^{-13}$ & $1.57$ & 303 \\ 
J0835\textminus4510 & 22.37 & 0.04 & $-3.22 \cdot 10^{-11}$ & $8.51 \cdot 10^{-14}$ & $0.89$ & 35 \\ 
J0940\textminus5428 & 22.84 & 0.05 & $-8.56 \cdot 10^{-12}$ & $2.50 \cdot 10^{-14}$ & $0.91$ & 11 \\ 
J1028\textminus5819 AG & 21.88 & 0.04 & $-3.86 \cdot 10^{-12}$ & $3.56 \cdot 10^{-14}$ & $0.33$ & 3 \\ 
J1028\textminus5819 BG & 21.88 & 0.04 & $-3.86 \cdot 10^{-12}$ & $2.63 \cdot 10^{-14}$ & $0.54$ & 5 \\ 
J1105\textminus6107 & 31.64 & 0.06 & $-7.94 \cdot 10^{-12}$ & $2.00 \cdot 10^{-14}$ & $1.26$ & 9 \\ 
J1112\textminus6103 & 30.78 & 0.06 & $-1.49 \cdot 10^{-11}$ & $3.50 \cdot 10^{-14}$ & $1.23$ & 15 \\ 
J1300+1240 & 321.62 & 0.64 & $-5.91 \cdot 10^{-15}$ & $5.00 \cdot 10^{-15}$ & $12.86$ & 3 \\ 
J1302\textminus6350 & 41.87 & 0.08 & $-2.00 \cdot 10^{-12}$ & $5.00 \cdot 10^{-15}$ & $1.67$ & 3 \\ 
J1400\textminus6325 & 64.12 & 0.13 & $-8.00 \cdot 10^{-11}$ & $1.65 \cdot 10^{-13}$ & $2.56$ & 67 \\ 
J1410\textminus6132 & 39.95 & 0.08 & $-2.52 \cdot 10^{-11}$ & $7.01 \cdot 10^{-14}$ & $1.60$ & 29 \\ 
J1420\textminus6048 & 29.32 & 0.06 & $-3.57 \cdot 10^{-11}$ & $1.00 \cdot 10^{-13}$ & $1.17$ & 41 \\ 
J1524\textminus5625 & 25.56 & 0.05 & $-1.27 \cdot 10^{-11}$ & $3.00 \cdot 10^{-14}$ & $1.02$ & 13 \\ 
J1531\textminus5610 & 23.75 & 0.05 & $-3.88 \cdot 10^{-12}$ & $1.50 \cdot 10^{-14}$ & $0.95$ & 7 \\ 
J1617\textminus5055 & 28.80 & 0.06 & $-5.62 \cdot 10^{-11}$ & $1.15 \cdot 10^{-13}$ & $1.15$ & 47 \\ 
J1718\textminus3825 BG & 26.78 & 0.05 & $-4.72 \cdot 10^{-12}$ & $1.72 \cdot 10^{-14}$ & $0.82$ & 5 \\ 
J1747\textminus2809 & 38.32 & 0.08 & $-1.14 \cdot 10^{-10}$ & $2.35 \cdot 10^{-13}$ & $1.53$ & 95 \\ 
J1747\textminus2958 & 20.23 & 0.04 & $-1.25 \cdot 10^{-11}$ & $3.00 \cdot 10^{-14}$ & $0.81$ & 13 \\ 
J1809\textminus1917 & 24.17 & 0.05 & $-7.44 \cdot 10^{-12}$ & $2.00 \cdot 10^{-14}$ & $0.97$ & 9 \\ 
J1811\textminus1925 & 30.91 & 0.06 & $-2.10 \cdot 10^{-11}$ & $4.50 \cdot 10^{-14}$ & $1.23$ & 19 \\ 
J1813\textminus1246 & 41.60 & 0.08 & $-1.52 \cdot 10^{-11}$ & $3.50 \cdot 10^{-14}$ & $1.66$ & 15 \\ 
J1813\textminus1749 & 44.71 & 0.09 & $-1.27 \cdot 10^{-10}$ & $2.60 \cdot 10^{-13}$ & $1.79$ & 105 \\ 
J1831\textminus0952 & 29.73 & 0.06 & $-3.67 \cdot 10^{-12}$ & $1.00 \cdot 10^{-14}$ & $1.19$ & 5 \\ 
J1833\textminus1034 & 32.29 & 0.06 & $-1.05 \cdot 10^{-10}$ & $2.15 \cdot 10^{-13}$ & $1.29$ & 87 \\ 
J1838\textminus0655 & 28.36 & 0.06 & $-1.99 \cdot 10^{-11}$ & $5.51 \cdot 10^{-14}$ & $1.13$ & 23 \\ 
J1913+1011 & 55.69 & 0.11 & $-5.25 \cdot 10^{-12}$ & $1.50 \cdot 10^{-14}$ & $2.23$ & 7 \\ 
J1952+3252 & 50.59 & 0.10 & $-7.48 \cdot 10^{-12}$ & $2.00 \cdot 10^{-14}$ & $2.02$ & 9 \\ 
J2022+3842 & 41.16 & 0.08 & $-7.30 \cdot 10^{-11}$ & $1.50 \cdot 10^{-13}$ & $1.64$ & 61 \\ 
J2043+2740 & 20.80 & 0.04 & $-2.75 \cdot 10^{-13}$ & $5.00 \cdot 10^{-15}$ & $0.83$ & 3 \\ 
J2124\textminus3358 & 405.59 & 0.81 & $-16.92 \cdot 10^{-16}$ & $5.00 \cdot 10^{-15}$ & $16.21$ & 3 \\ 
J2229+6114 & 38.71 & 0.08 & $-5.84 \cdot 10^{-11}$ & $1.20 \cdot 10^{-13}$ & $1.55$ & 49 \\

\end{tabular}
\end{table*}

\section{Results \label{sec:4}}
The search has produced a total of 49 outliers for 15 of the 33 targets. Every outlier underwent a chain of follow-up steps aimed to test its nature, namely: \textit{i)} check for  the presence of known instrumental noise lines,  \textit{ii)} comparing the SNR GW amplitude estimation among several detectors and  \textit{iii)} studying the outlier significance with software injections. The outliers are given in Tab. \ref{tab:out_recap} together with the step of the follow-up where we excluded them. 

\begin{table}
\caption{\label{tab:out_recap}This table summarizes the outliers found in the O2 narrowband search. The first column reports the name of the pulsar for which we have found outliers. The second column gives the central frequency of the pulsar search band and the third column the p-value of the least significant outlier. The last column reports the step of the follow-up in which we have vetoed the outliers. For a description of the follow-up steps refer to the main text.}
\begin{tabular}{l|cccc}
\textbf{Name}& $f$& num cand. & p-value & Step \\ \hline
J1105\textminus6107 & 31.64 & 16\footnote{most vetoed since they are close to  the comb line of  0.987925~Hz comb in LLO and comb line of 2.109223~Hz in LHO.} & $4.23 \times 10^{-4}$ & \rom{1},\rom{2}\\
J1112\textminus6103 & 30.78& 1\footnote{Various unidentified  lines around 35.51~Hz.}  & $1.83 \times 10^{-4}$&\rom{2}\\
J1300+1240 & 321.62 & 1 &  $7.80 \times 10^{-4}$  & \rom{3}\\
J1302\textminus6350 & 41.87 & 4\footnote{Unidentified noise disturbance in LHO at 41.8838~Hz.} &$7.79 \times 10^{-4}$&\rom{2}\\
%J1410\textminus6132 & 39.95 & 1 & $1.03 \times 10^{-5}$ &\rom{2} \\
J1420\textminus6048 & 29.32 & 11\footnote{Comb of 1.945501~Hz in LHO.} &$9.82 \times 10^{-4}$ & \rom{1},\rom{2} \\
J1531\textminus5610 & 23.75 & 1 & $4.65\times 10^{-4}$ & \rom{2} \\
J1617\textminus5055 & 28.80 & 2 & $7.80 \times 10^{-4}$ & \rom{2},\rom{3} \\
J1747\textminus2809 & 38.32 & 1 & $9.68 \times 10^{-4}$ &\rom{2} \\
J1811\textminus1925 & 30.91 & 1 &$3.30 \times 10^{-4}$&  \rom{2} \\
J1813\textminus1246 & 41.60 & 2\footnote{Unidentified broad line disturbance at 41.654-41.660~Hz.}  &$6.73 \times 10^{-4}$& \rom{2},\rom{3}\\
J1831\textminus0952 & 29.73 & 1 &$2.15 \times 10^{-4}$&\rom{2}\\
J1833\textminus1034 & 32.29 & 1 &$9.33 \times 10^{-4}$& \rom{2} \\
%J1838\textminus0655 & 28.36 & 2 &$2.95 \times 10^{-4}$& \rom{2},\rom{3} \\
J1952+3252 & 50.59 & 4\footnote{comb of 2.109223~Hz in LHO, comb of 1.9455045~Hz in LHO, comb of 1.945437~Hz in LHO.} &$4.48 \times 10^{-4}$& \rom{1},\rom{2} \\
J2124\textminus3358 & 405.59 & 2\footnote{Comb of 0.9967943~Hz in LLO.} &$5.61 \times 10^{-4}$&\rom{1},\rom{3}\\
J2229+6114 & 38.71 & 1 &$9.66 \times 10^{-4}$& \rom{2}\\
\end{tabular}
\end{table}
The narrow-band search carried out in the past on O1 data \citep{2017PhRvD..96l2006A} produced two interesting outliers for J0835\textminus4510 and 1833\textminus1034. In order to confirm or reject them, the data from the first four months of  O2 (available with calibration version C01 at the time) were used and no evidence for a signal was found. The full O2 analysis discussed in this paper confirms those findings. No outlier has been found for J0835\textminus4510, while an outlier has been found for J1833\textminus1034, at a slightly different frequency which however, as discussed in the next section, has been vetoed.

\subsection{Outliers follow-up \label{sec:outliers}}
The first step of the follow-up was to check if a known instrumental noise line was present in one of the two detectors \cite{2018PhRvD..97h2002C}. This ruled out most of the candidates for the pulsars J1105\textminus6107 and J2121\textminus3358, see Appendix C for more details.

The second step of the follow-up was to study the evolution of the recovered signal-to-noise ratio (SNR) and amplitude $h_0$ with respect to the fraction of data samples that we are integrating. We expect the SNR to increase as the square root of the integration time and the amplitude $h_0$ to be nearly constant. We have performed this type of test in a LHO, LLO and joint search for different integration times, checking if the SNR and $h_0$ estimation were compatible across the different cases.

Many outliers at frequencies $<100$~Hz have been classified as LHO disturbances, since they have been observed only in LHO (see Appendix C).  Some of these are in proximity of unidentified noise lines (lines which are confidently classified as detector disturbances, but whose origin is unknown). That is the case of the outliers from J1112\textminus6103, J1302\textminus6350 and J1813\textminus1246. Other outliers at low frequency were not in proximity of unidentified noise lines but have been vetoed as the signal-to-noise ratio is bigger than $8$ only in LHO data, which has a sensitivity 2 to 3 times worse than LLO, thus being incompatible with a true CW signal.

Only 3 outliers survived up to the third step of the follow-up, namely from pulsars J1300+1240, J1617\textminus5055 and J2124\textminus3358. For all these pulsars we cannot approach  the theoretical spin-down limit with our current search sensitivity, and this is a strong hint for the noise origin of these outliers. The last step of the follow-up  consisted in studying the SNR and recovered CW amplitude $h_0$  with software injections with an amplitude $h_0$ fixed to that estimated for the outlier. The evolution of the SNR and $h_0$ for the outlier is then compared to the distributions derived from the injections. If they are compatible among the three different analyses, LHO, LLO and joint combination, the outlier is subject to more dedicated studies.  
The two remaining outliers for the millisecond pulsars were ruled out since they were present in just one detector, while the injections predicted that they would be visible in both the detectors. 
The J1617\textminus5055 remaining  outlier were also ruled out, as the injections show that they were likely driven by an LHO disturbance. Refer to Appendix C for more details on the last steps of this follow-up.

\subsection{Upper limits}
Since there was no evidence for the presence of a CW signal, we have computed upper limits (ULs) on the CW amplitude $h_0$. The ULs have been produced using the  same procedure as in the O1 narrow-band search \cite{2017PhRvD..96l2006A}, which consists in injecting non-overlapping GW signals with fixed amplitude $h_0$ in data every $10^{-4}$~Hz intervals. When the 95\% of injections produce a value of the detection statistic higher than the one used for the outlier selection, we set the upper-limit to the injected amplitude value.  

Fig. \ref{fig:overall_plot} shows the median value of the UL for each of the 33 targets. The ULs are driven at lower frequencies by LLO sensitivity, since it is the most sensitive detector in that frequency region. On the other hand, at higher frequencies the ULs lie  close to the sensitivity of the two detectors, which are indeed  similar.

Tab. \ref{tab:O2_UL} summarizes our results for the O2 narrow-band search. The table reports the median value of the UL on the strain amplitude $h_0$ and the corresponding ellipticity, computed using Eq. \eqref{eq:eps_sd_limit}. We consider the spin-down limit surpassed for a given pulsar, if the ULs are lower than the spin-down limit over the entire frequency band. 

The most stringent ULs have been set for the 3 pulsars J0537\textminus6910, J1300+1240 and J2124\textminus3358 and are of the order of $5.5 \times 10^{-26}$ which, however, are above the spin-down limit. The lowest ellipticity UL has been set for J1300+1240, of about $3.3 \times 10^{-7}$.
We have been able to surpass the spin-down limit for the pulsars: J0205+6449, J0534+2200 (Crab), J0835\textminus4510 (Vela), J1400\textminus6325, J1813\textminus1246 (assuming the lower bound for the distance), J1813\textminus1749, J1833\textminus1034 and J2229+6114.
For J0940\textminus5428, while the median value of the UL is below the spin-down limit, a small fraction of the individual results are above. 
For J1747\textminus2809 and J1952+3252 we are close to surpassing the spin-down limit\footnote{Excluding a frequency band heavily contaminated by noise.}, see Tab. \ref{tab:O2_UL}.
For all the pulsars for which we have surpassed the spin-down limit, we have computed the upper limit on the ratio of the GW to the rotational energy loss. The lower ULs on the GW energy loss are for J0534+2200 and J1400\textminus6325, corresponding to a fraction of about 0.8\%.
The lowest ULs on the GW amplitude and ellipticity among the pulsars for which we have surpassed the spin-down limit are, respectively, $8.29 \times 10^{-26}$  and $1.78 \times 10^{-5}$, for J1400\textminus6325.
For a canonical pulsar with a radius of about 10 km, this number would correspond to a maximum surface deformation of about $5$ cm.

For the remaining 22 targets we were not able to surpass the spin-down limit. Tab. \ref{tab:O2_UL} roughly suggests to us that an improvement in sensitivity of a factor 3 is needed for most of the low-frequency pulsars. It must be considered, however, that the spin-down limits have been computed assuming a canonical value for the moment of inertia of $10^{38} \mathrm{kg \, m^2}$. In fact, it could be significantly larger, depending on NS equation of state, up to $\sim 3\times 10^{38} \mathrm{kg \, m^2}$, implying a spin-down limit $\sim \sqrt(3)$ times larger. 

\begin{table*}[h]
\caption{\label{tab:O2_UL}Upper limits summary table. First column: pulsar name. Second and third columns: median of the  95\% confidence level UL on the GW amplitude $h_0$ and corresponding ellipticity $\epsilon$. Fourth column: surface deformation corresponding to the median ellipticity for a NS with radius of 10 km \cite{2013PhRvD..88d4016J}. Fifth column: ratio between the median UL and the spin-down limit. Sixth column: ratio between the median UL on the GW and rotational energy losses. Last column: minimum and maximum ratio between the ULs and the theoretical spin-down limit over the analyzed frequency/spindown region. All the entries that use information on the astrophysical distance  also include the corresponding uncertainty at $1\sigma$ confidence level.}
\begin{tabular}{l|ccccc|c}
\textbf{Name}& $\braket{h}_{\rm UL}$& $\braket{\epsilon}_{\rm UL}$& $r_\epsilon$[cm] &$\braket{h}_{\rm UL}/h_{\rm sd}$&$\braket{\dot{E}_{\rm UL}}/\dot{E}_{\rm sd}$& ${\rm min}_{\rm nb}[\braket{h}_{\rm UL}/h_{\rm sd}] \,-\,{\rm max}_{\rm nb}[\braket{h}_{\rm UL}/h_{\rm sd}]$\\ \hline

J0205+6449 AG & $3.87 \cdot 10^{-25}$ & $(7.9 \pm 1.2)\cdot 10^{-4}$ & $197.9$ & $0.56 \pm 0.08$ & $0.3$ & $0.48^{+0.07}_{-0.07} -0.67^{+0.10}_{-0.10}$ \\ 
J0205+6449 BG & $3.19 \cdot 10^{-25}$ & $(6.5 \pm 1.0)\cdot 10^{-4}$ & $163.1$ & $0.46 \pm 0.07$ & $0.2$ & $0.31^{+0.05}_{-0.05} -0.58^{+0.09}_{-0.09}$ \\ 
J0534+2200 AG & $1.31 \cdot 10^{-25}$ & $(7.1 \pm 1.8)\cdot 10^{-5}$ & $17.4$ & $0.09 \pm 0.02$ & $0.008$ & $0.07^{+0.02}_{-0.02} -0.11^{+0.03}_{-0.03}$ \\ 
J0534+2200 BG & $1.64 \cdot 10^{-25}$ & $(8.8 \pm 2.2)\cdot 10^{-5}$ & $21.7$ & $0.11 \pm 0.03$ & $0.01$ & $0.09^{+0.02}_{-0.02} -0.14^{+0.03}_{-0.03}$ \\ 
J0537\textminus6910 & $5.59 \cdot 10^{-26}$ & $(1.7 \pm 0.01)\cdot 10^{-4}$ & -&$1.92 \pm 0.01$ & -&$1.13^{+0.00}_{-0.00} -2.25^{+0.01}_{-0.01}$ \\ 
J0540\textminus6919 & $1.47 \cdot 10^{-25}$ & $(4.43 \pm 0.02)\cdot 10^{-3}$ & -&$2.95 \pm 0.01$ & -&$1.83^{+0.01}_{-0.01} -3.47^{+0.02}_{-0.02}$ \\ 
J0835\textminus4510 & $8.82 \cdot 10^{-25}$ & $(4.7 \pm 0.4)\cdot 10^{-4}$ & $116.8$ & $0.26 \pm 0.02$ & $0.07$ & $0.14^{+0.01}_{-0.01} -0.31^{+0.02}_{-0.02}$ \\ 
J0940\textminus5428 & $8.55 \cdot 10^{-25}$ & $(5.9 \pm 2.3)\cdot 10^{-4}$ & $147.6$ & $0.7 \pm 0.3$ & $0.5$ & $0.4^{+0.2}_{-0.2} -0.8^{+0.4}_{-0.4}$ \\ 
J1028\textminus5819 AG& $1.18 \cdot 10^{-24}$ & $(3.3 \pm 1.3)\cdot 10^{-3}$ & -&$5.0 \pm 2.0$ & -&$4.2^{+1.7}_{-1.7} -6.0^{+2.3}_{-2.3}$ \\ 
J1028\textminus5819 BG& $1.37 \cdot 10^{-24}$ & $(3.8 \pm 1.5)\cdot 10^{-3}$ & -&$5.7 \pm 2.3$ & -&$4.3^{+1.7}_{-1.7} -7.0^{+2.7}_{-2.7}$ \\ 
J1105\textminus6107 & $2.20 \cdot 10^{-25}$ & $(5.0 \pm 2.0)\cdot 10^{-4}$ & $123.0$ & $1.3 \pm 0.6$ & $1.7$ & $0.68^{+0.3}_{-0.3} -1.88^{+0.8}_{-0.8}$ \\ 
J1112\textminus6103 & $2.48 \cdot 10^{-25}$ & $(1.1 \pm 0.5)\cdot 10^{-3}$ & -&$2.0 \pm 0.8$ & -&$1.1^{+0.5}_{-0.5} -2.5^{+1.0}_{-1.0}$ \\ 
J1300+1240 & $5.60 \cdot 10^{-26}$ & $(3.3 \pm 0.8)\cdot 10^{-7}$ & -&$10.5 \pm 2.5$ & -&$6.3^{+1.5}_{-1.5} -13.1^{+3.1}_{-3.1}$ \\ 
J1302\textminus6350 & $1.22 \cdot 10^{-25}$ & $(1.5 \pm 0.6)\cdot 10^{-4}$ & $38.0$ & $1.6 \pm 0.7$ & $2.6$ & $0.7^{+0.3}_{-0.3} -1.9^{+0.8}_{-0.8}$ \\ 
J1400\textminus6325 & $8.57 \cdot 10^{-26}$ & $(1.8 \pm 0.6)\cdot 10^{-5}$ & $4.4$ & $0.09 \pm 0.03$ & $0.008$ & $0.05^{+0.01}_{-0.01} -0.10^{+0.03}_{-0.03}$ \\ 
J1410\textminus6132 & $1.33 \cdot 10^{-25}$ & $(1.1 \pm 0.5)\cdot 10^{-3}$ & -&$2.8 \pm 1.1$ & -&$1.4^{+0.6}_{-0.6} -3.5^{+1.4}_{-1.4}$ \\ 
J1420\textminus6048 & $2.75 \cdot 10^{-25}$ & $(1.7 \pm 0.7)\cdot 10^{-3}$ & $426.8$ & $1.7 \pm 0.7$ & $3.1$ & $0.9^{+0.4}_{-0.4} -2.2^{+0.9}_{-0.9}$ \\ 
J1524\textminus5625 & $5.03 \cdot 10^{-25}$ & $(2.5 \pm 1.0)\cdot 10^{-3}$ & -&$3.0 \pm 1.2$ & -&$1.7^{+0.7}_{-0.7} -3.7^{+1.5}_{-1.5}$ \\ 
J1531\textminus5610 & $7.51 \cdot 10^{-25}$ & $(3.6 \pm 1.4)\cdot 10^{-3}$ & -&$6.5 \pm 2.6$ & -&$3.7^{+1.5}_{-1.5} -7.7^{+3.1}_{-3.1}$ \\ 
J1617\textminus5055 & $3.41 \cdot 10^{-25}$ & $(1.8 \pm 0.6)\cdot 10^{-3}$ & $461.0$ & $1.5 \pm 0.6$ & $2.1$ & $0.8^{+0.3}_{-0.3} -1.8^{+0.8}_{-0.8}$ \\ 
J1718\textminus3825 BG& $3.88 \cdot 10^{-25}$ & $(1.8 \pm 0.7)\cdot 10^{-3}$ & -&$4.0 \pm 1.6$ & -&$2.5^{+1.0}_{-1.0} -4.8^{+2.0}_{-2.0}$ \\ 
J1747\textminus2809 & $1.43 \cdot 10^{-25}$ & $(7.5 \pm 2.9)\cdot 10^{-4}$ & $188.1$ & $0.8 \pm 0.4$ & $0.6$ & $0.5^{+0.2}_{-0.2} -1.0^{+0.4}_{-0.4}$ \\ 
J1747\textminus2958 & $1.35 \cdot 10^{-24}$ & $(7.9 \pm 3.1)\cdot 10^{-3}$ & -&$5.4 \pm 2.1$ & -&$3.2^{+1.3}_{-1.3} -6.7^{+2.6}_{-2.6}$ \\ 
J1809\textminus1917 & $6.95 \cdot 10^{-25}$ & $(3.7 \pm 1.5)\cdot 10^{-3}$ & -&$5.1 \pm 2.0$ & -&$3.1^{+1.2}_{-1.2} -6.2^{+2.4}_{-2.4}$ \\ 
J1811\textminus1925 & $2.53 \cdot 10^{-25}$ & $(1.2 \pm 0.5)\cdot 10^{-3}$ & -&$1.9 \pm 0.8$ & -&$1.3^{+0.5}_{-0.5} -2.3^{+0.9}_{-0.9}$ \\ 
J1813\textminus1246 & $1.23 \cdot 10^{-25}$ & $\leq 7\cdot 10^{-4}$ & $42.0$ & $\geq 0.7 $ & $\geq 0.5$ & $\geq (0.4-0.8)$ \\ 
J1813\textminus1749 & $1.16 \cdot 10^{-25}$ & $(2.6 \pm 0.5)\cdot 10^{-4}$ & $64.5$ & $0.40 \pm 0.07$ & $0.2$ & $0.25^{+0.04}_{-0.04} -0.49^{+0.08}_{-0.08}$ \\ 
J1831\textminus0952 & $2.56 \cdot 10^{-25}$ & $(1.0 \pm 0.4)\cdot 10^{-3}$ & -&$3.3 \pm 1.3$ & -&$2.1^{+0.9}_{-0.9} -4.2^{+1.7}_{-1.7}$ \\ 
J1833\textminus1034 & $1.96 \cdot 10^{-25}$ & $(7.3 \pm 0.6)\cdot 10^{-4}$ & $182.5$ & $0.55 \pm 0.04$ & $0.3$ & $0.35^{+0.03}_{-0.03} -0.71^{+0.05}_{-0.05}$ \\ 
J1838\textminus0655 & $3.03 \cdot 10^{-25}$ & $(2.4 \pm 0.4)\cdot 10^{-3}$ & -&$3.0 \pm 0.4$ & -&$1.8^{+0.3}_{-0.3} -3.6^{+0.5}_{-0.5}$ \\ 
J1913+1011 & $1.02 \cdot 10^{-25}$ & $(1.4 \pm 0.6)\cdot 10^{-4}$ & -&$1.9 \pm 0.8$ & -&$1.1^{+0.5}_{-0.5} -2.31^{+0.9}_{-0.9}$ \\ 
J1952+3252 & $9.09 \cdot 10^{-26}$ & $(1.0 \pm 0.7)\cdot 10^{-4}$ & $25.2$ & $0.9 \pm 0.6$ & $0.8$ & $0.5^{+0.4}_{-0.4} -1.1^{+0.8}_{-0.8}$ \\ 
J2022+3842 & $1.32 \cdot 10^{-25}$ & $(7.4 \pm 1.5)\cdot 10^{-4}$ & $184.0$ & $1.2 \pm 0.3$ & $1.4$ & $0.7^{+0.2}_{-0.2} -1.5^{+0.3}_{-0.3}$ \\ 
J2043+2740 & $1.12 \cdot 10^{-24}$ & $(3.6 \pm 1.4)\cdot 10^{-3}$ & -&$17.8 \pm 7.0$ & -&$10.3^{+4.0}_{-4.0} -21.42^{+9.0}_{-9.0}$ \\ 
J2124\textminus3358 & $5.97 \cdot 10^{-26}$ & $(1.3 \pm 0.3)\cdot 10^{-7}$ & -&$14.0 \pm 3.3$ & -&$7.3^{+1.8}_{-1.8} -17.4^{+4.2}_{-4.2}$ \\ 
J2229+6114 & $1.39 \cdot 10^{-25}$ & $(2.7 \pm 1.8)\cdot 10^{-4}$ & $65.8$ & $0.4 \pm 0.3$ & $0.2$ & $0.3^{+0.2}_{-0.2} -0.5^{+0.4}_{-0.4}$ \\ 

\end{tabular}
\end{table*}
\section{Conclusion \label{sec:5}}
Overall, the narrow-band search over O2 data has brought an improvement with respect to previous searches in terms of ULs.
On the other hand, ULs are similar to those found in O1 for pulsars with rotation frequency below $30$~Hz. For instance the UL on the Vela pulsar (around $22$~Hz) has improved by 10\%, while the UL on J0205+6449\footnote{Please note that the spin-down limit of this pulsar has been computed using two different distance in O1 and O2. For O1 we used 2.0 kpc \cite{2013A&A...560A..18K} while for O2 the nominal ephemeris value  was 3.2 kpc.} has improved by about 22\%. On the other hand for pulsars with expected GW frequencies $>30$~Hz the UL is improved even by a factor 2.
The UL on J0534+2200 did not improve, since in O2 we split the analysis in two different chunks due to the presence of the glitch. For this reason the UL, both before and after the glitch, is comparable with the one found in O1 analysis. We have also been able to surpass the spin-down limit for two pulsars that were not analyzed in O1, J0940\textminus5428, J1747\textminus2809.

We are still not able to surpass the spin-down limit for the millisecond pulsars and for low frequency pulsars with spin-down below $\sim 10^{-12}$~Hz/s. However, we are able to surpass the spin-down limit for low frequency and high energetic pulsars (such as Crab or J1833\textminus1034) or for low frequency pulsars that are close to the Earth.

\section*{Acknowledgments}
%The authors are thankful to the anonymous referees for the help to improve this paper.
The authors gratefully acknowledge the support of the United States National Science Foundation
(NSF) for the construction and operation of the LIGO Laboratory and Advanced LIGO as well as the
Science and Technology Facilities Council (STFC) of the United Kingdom, the Max-Planck-Society
(MPS), and the State of Niedersachsen/Germany for support of the construction of Advanced LIGO  and
construction and operation of the GEO600 detector.  Additional support for Advanced LIGO was
provided by the Australian Research Council. The authors gratefully acknowledge the Italian Istituto
Nazionale di Fisica Nucleare (INFN), the French Centre National de la Recherche Scientifique (CNRS)
and the Foundation for Fundamental Research on Matter supported by the Netherlands Organisation for
Scientific Research, for the construction and operation of the Virgo detector and the creation and
support  of the EGO consortium. The authors also gratefully acknowledge research support from these
agencies as well as by the Council of Scientific and Industrial Research of India, the Department of
Science and Technology, India, the Science \& Engineering Research Board (SERB), India, the Ministry
of Human Resource Development, India, the Spanish Agencia Estatal de Investigaci\'on, the
Vicepresid\`encia i Conselleria d'Innovaci\'o, Recerca i Turisme and the Conselleria d'Educaci\'o i
Universitat del Govern de les Illes Balears, the Conselleria d'Educaci\'o, Investigaci\'o, Cultura i
Esport de la Generalitat Valenciana, the National Science Centre of Poland, the Swiss National
Science Foundation (SNSF), the Russian Foundation for Basic Research, the Russian Science
Foundation, the European Commission, the European Regional Development Funds (ERDF), the Royal
Society, the Scottish Funding Council, the Scottish Universities Physics Alliance, the Hungarian
Scientific Research Fund (OTKA), the Lyon Institute of Origins (LIO), the National Research,
Development and Innovation Office Hungary (NKFI),  the National Research Foundation of Korea,
Industry Canada and the Province of Ontario through the Ministry of Economic Development and
Innovation, the Natural Science and Engineering Research Council Canada, the Canadian Institute for
Advanced Research, the Brazilian Ministry of Science, Technology, Innovations, and Communications,
the International Center for Theoretical Physics South American Institute for Fundamental Research
(ICTP-SAIFR),  the Research Grants Council of Hong Kong, the National Natural Science Foundation of
China (NSFC), the Leverhulme Trust, the Research Corporation, the Ministry of Science and Technology
(MOST), Taiwan and the Kavli Foundation. The authors gratefully acknowledge the support of the NSF,
STFC, MPS, INFN, CNRS and the State of Niedersachsen/Germany for provision of computational
resources.  Work at Naval Research Laboratory (NRL) is supported by NASA.
The authors acknowledge the anonymous referees for helping to improve this paper.
This work has been assigned LIGO document number \dcc.

\appendix
\section{Validation with hardware injections}
Hardware injections are simulated signals in LIGO-Virgo data for testing purposes. These artificial signals  are  injected  by a control system which acts on the mirror and simulate a CW signal. The Hardware injections are continuously monitored and their  injected parameters are known.
In order to validate the efficiency of the pipeline used in this paper, we have looked for 4 Hardware injections in LIGO data studying the accuracy of the recovered parameters. 
We define the relative error on the CW amplitude recovery as $\epsilon_{h_0} = 1-h_{0}^{\rm esti}/h_{0}^{\rm inj}$, where $h_{0}^{\rm inj}$ is the injected CW amplitude and $h_{0}^{\rm esti}$ is the recovered value. Whereas we define the relative error on the angular parameters $\psi, \eta$ as $\epsilon_\psi=|\psi^{\rm inj}-\psi^{\rm esti}|/90 {\rm  \, deg}$ and  $\epsilon_\eta=|\eta^{\rm inj}-\eta^{\rm esti}|/2$. Tab. \ref{tab:HI_unc} reports the errors on the parameter estimation for the validation tests performed with the O2 hardware injections.

\begin{table}
\caption{\label{tab:HI_unc} Accuracy of the parameter estimation for the O2 hardware injections. The first three columns report the name, frequency and spin-down of the hardware injections (reference time at Dec 1st 2017 UTC 00:00:00). The last three columns report the relative errors in percentage for the parameter estimation. The relative errors are defined in the text.}
\begin{tabular}{l|ccccc}
\textbf{Name}& $f_{\rm gw}$ [Hz]&$\dot{f}_{\rm gw}$ [Hz/s] & $\epsilon_{h_0}$ & $\epsilon_{\eta}$ & $\epsilon_{\psi}$ \\ \hline
Pulsar 2 & 575.16 & $-1.37 \cdot 10^{-13}$ &  6\% & 0.3\% & - \\
Pulsar 3 & 108.86 & $-1.46 \cdot 10^{-17}$ &  0.01\% & 0.3\% & 2\% \\
Pulsar 5 & 52.81 & $-4.03 \cdot 10^{-18}$ &  3\% & 0.07\%  & 1\% \\
Pulsar 8 & 190.46 & $-8.65 \cdot 10^{-9}$ &  8\% & 0.03\%  &  0.07\%  \\

\end{tabular}
\end{table}

\section{validation of the threshold }
\label{app:thr}
\begin{figure}
\centering
\includegraphics[width=0.5\textwidth]{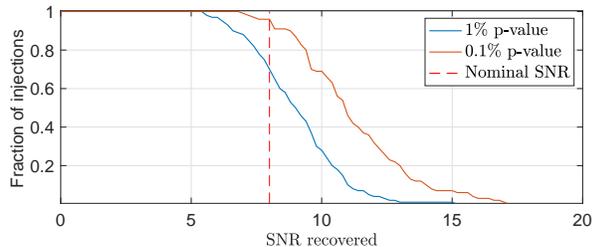}
\caption{\textit{Vertical axis:} fraction of injections recovered with an SNR equal or higher than the one indicated on the \textit{horizontal axis}. The different line colors indicate a set of software injections that would produce an outlier at $1\%$ and $0.1\%$ according to the evaluation of the noise-only distribution of the detection statistic. The red-dashed vertical line indicates the SNR=8 threshold that is commonly used to distinguish the signal from the noise.}
\label{fig:effect_of_the_thr}
\end{figure}

The narrow-band search is based on the 5-vector method \cite{2010CQGra..27s4016A}, that was implemented originally for {\it targeted searches}. In that context just one template is explored for each detector, and an overall threshold on the p-value of, say, 1\% for the candidate selection is sufficient to efficiently recover 95\% of injected signals with SNR=8. 
However, in narrow-band searches we are exploring a large number of templates in a frequency region of about $0.04$~Hz or more, using two detectors that have different data quality, i.e. different level of noise and duty cycle. The threshold in this case is computed by using as noise background the values of the statistic excluded from the local maxima selection and then extrapolating the long tails of the distribution. By definition, these excluded points are representative of the noise level in the given frequency bands. This means that, if the noise level in the  $10^{-4}$~Hz wide frequency sub-band that we are analyzing is slightly higher than the noise level in the overall frequency region from which we are generating noise backgrounds, then close-to-threshold outliers will occur. These close-to-threshold outliers may be not completely distinguishable from  the actual noise.
As an example, we have generated 200 software injections with amplitude $h_0$ fixed to the one that generated a $1\%$ p-value outlier in the post-glitch analysis of pulsar J0534+2200. We have estimated the recovered signal-to-noise ratio of the injections by integrating coherently more and more data from LHO and LLO. If the injections are distinguishable from the noise, we expect  95\% of  the injections to have a recovered signal-to-noise ratio greater than $8$. However, it is shown by Fig. \ref{fig:effect_of_the_thr} this is not the case. 
For a full coherent LHO-LLO search, the distribution of the recovered SNR is below 8. We have also performed the same test by injecting fake signals with an amplitude $h_0$ that would correspond to a $0.1\%$ outlier. In this case, as shown in Fig. \ref{fig:effect_of_the_thr}, the recovered SNR of the injections is higher than 8, confirming that the 0.1\% p-value threshold represents a more conservative choice while recovering CW signals.

\section{Follow-up test cases \label{sec:appc}}
We report in this appendix some explanatory plots of the analysis steps used for outliers follow-up. The first step consisted in checking if a known noise line was present in the proximity of the outlier. We  considered an outlier  consistent with a known noise disturbance if it is found in a frequency region covered by the frequency variation of the noise line due to the Doppler and spin-down corrections.

\begin{figure}
\centering
\includegraphics[width=0.5\textwidth]{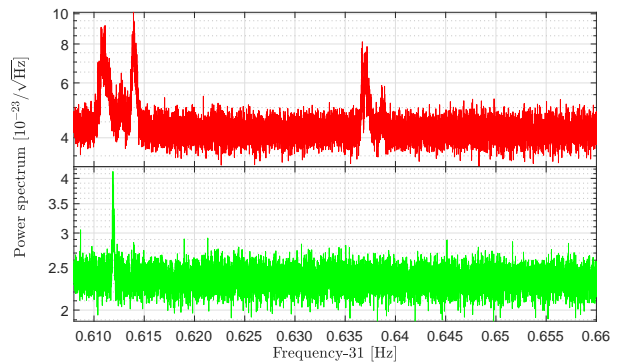}
\caption{\textit{Top}: LHO spectrum around the expected frequency of J1105\textminus6107. \textit{Bottom}: LLO spectrum around the expected signal frequency of J1105\textminus6107. In both the detectors, we see the contribution of various noise lines which are known comb with fundamental frequency 0.987925~Hz  in LLO and 2.109223~Hz in LHO.}
\label{fig:noise_J1105}
\end{figure}

\begin{figure}
\centering
\includegraphics[width=0.5\textwidth]{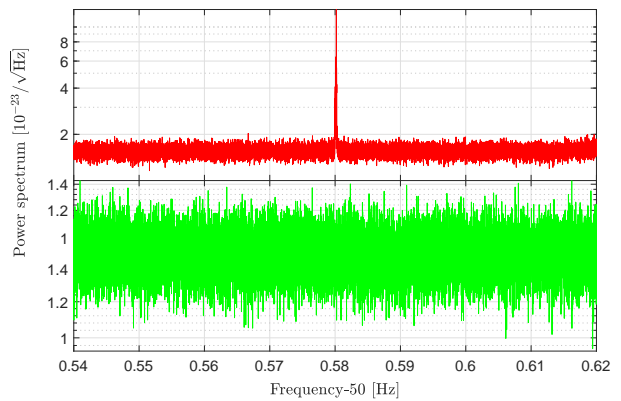}
\caption{\textit{Top}: LHO spectrum around the expected frequency of J1952+3252. \textit{Bottom}: LLO spectrum around the expected frequency of J1952+3252. In LHO we see the contribution of various noise lines due to combs with fundamental frequencies 2.109223~Hz, 1.9455045~Hz and 1.945437 in LHO.}
\label{fig:noise_J1952}
\end{figure}

Many of the outliers found in the case of the pulsar J1105\textminus6107 and J1952+3252 originated from vetoed combs in one or both of the detectors. Fig. \ref{fig:noise_J1105} and  Fig.  \ref{fig:noise_J1952} show the spectra of the time series obtained for J1105\textminus6107 and J1952+3252 outliers. In the first case,  noise combs pollute both LLO and LHO, while in the second case different noise combs contribute to the same noise disturbance at 50.58~Hz in LHO data.

\begin{figure}
\centering
\includegraphics[width=0.5\textwidth]{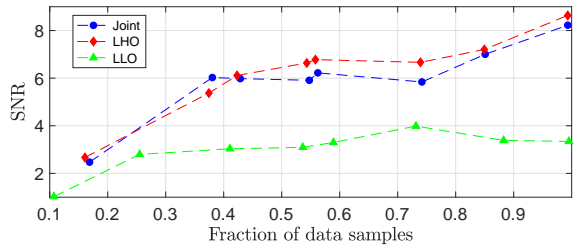}
\caption{Example of the first stage follow-up for one of the not  candidates of J1105+6107 that were not vetoed. The recovered SNR of the outlier is on the vertical axis while the horizontal axis indicates the fraction of data samples that we are integrating with the matched filter. The outlier is visible only in LHO and propagates to the joint analysis.}
\label{fig:example_first_stage_followup}
\end{figure}

The second step of the follow-up chain was to study the evolution of the recovered CW amplitude $h_0$ and  the recovered SNR of the outlier with respect to the integration time. In Fig. \ref{fig:example_first_stage_followup} we report the  recovered SNR for different integration times. In this frequency region, the LHO noise floor is about two times higher than the LLO noise floor. Hence in the presence of a reliable CW outlier, we would expect the recovered SNR to be higher in LLO and the joint analysis. As shown in Fig.  \ref{fig:example_first_stage_followup}, this is not the case and the outlier is probably due to an unknown noise disturbance in LHO.

\begin{figure*}
\centering
\hspace*{-5.0cm}
\includegraphics[width=1.5\textwidth]{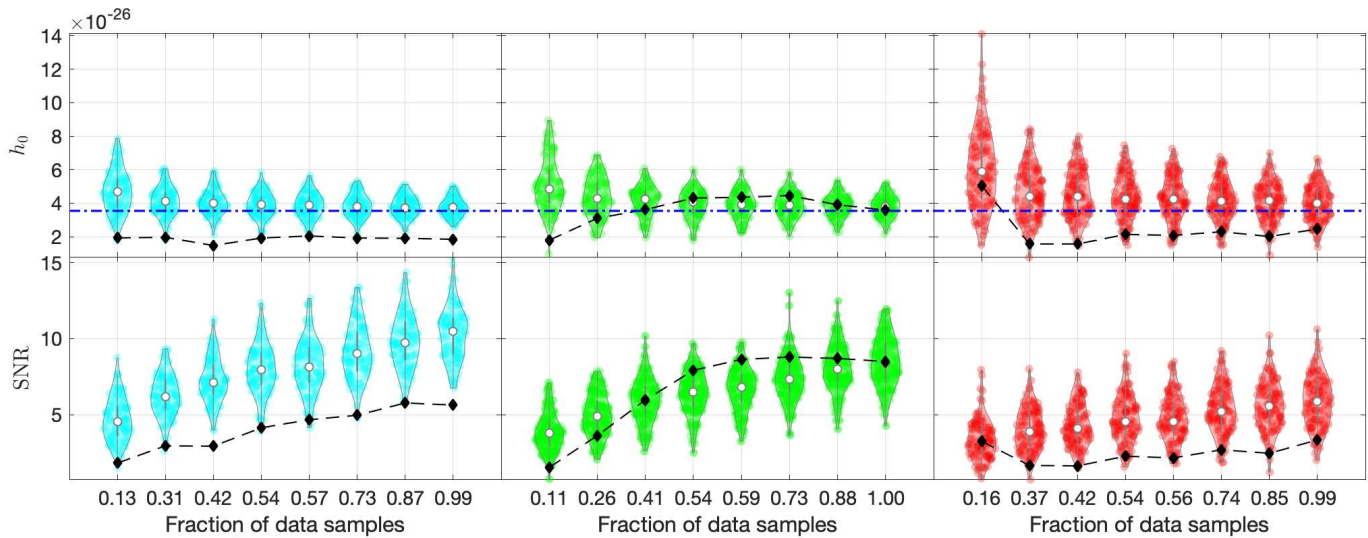}
\vspace*{-4.5cm}
\caption{These plots show the distribution of the recovered CW amplitude $h_0$ and SNR for 200 software injections in the frequency region around the outlier of the millisecond pulsar J1300+1240. The black dashed line indicates the observed estimator for the outlier. \textit{First and second rows of plots}: Recovered CW amplitude and SNR. \textit{First, second and third columns of plots:} Joint, LLO and LHO searches. }
\label{fig:IIstepfupexample}
\end{figure*}

The last step of the follow-up consisted in studying the noise properties with software injections around the candidates. The software injections had amplitude $h_0$ equal to the one recovered from the most sensitive search. This corresponds to LLO for most of the frequencies $<40$~Hz, while it is the joint search if the noise floor of the two detectors is comparable. The recovered distribution of the CW amplitude and SNR for the software injections is then plotted with respect to the integration time of  the analysis and compared with the recovered CW amplitude and SNR for the outlier. Fig. \ref{fig:IIstepfupexample} shows the distributions of the recovered SNR and CW amplitude for 200 software injections with an amplitude fixed at $h_0 =3.9 \times 10^{-26}$, which is the one recovered for the outlier of the millisecond pulsar J1300+1240 in the joint search. The software injections have a frequency at least $10^{-3}$~Hz away from the actual outlier, in such a way to not interfere with the outlier. From  Fig. \ref{fig:IIstepfupexample} we can see that the outlier seems to be compatible with the results of the software injections in LLO data, but on the other hand it is not compatible with the joint and LHO analysis. In this frequency region, the detectors noise floor is similar and we would expect comparable results for the LLO and LHO analysis.  The software injections  show that a signal with amplitude $h_0 \approx 3.9 \times 10^{-26}$ would be distinguishable from the noise in the joint search because the recovered SNR of the software injections with the same amplitude for a joint full coherent search is always higher than 7.5. On the contrary, in the joint search the SNR of the actual outlier (black dashed line) is  low and not compatible with the results of the software injections,  suggesting that the outlier is due to a unknown noise disturbance present in LLO.

\bibliographystyle{unsrt}
\bibliography{O2cw}

\begin{thebibliography}{10}

\bibitem{0264-9381-32-7-074001}
{The LIGO Scientific Collaboration} and J.~et~al. {Aasi}.
\newblock Advanced ligo.
\newblock {\em Classical and Quantum Gravity}, 32(7):074001, 2015.

\bibitem{PhysRevLett.116.131103}
{The LIGO Scientific Collaboration}, {the Virgo Collaboration}, and B.~P.
  et~al. {Abbott}.
\newblock Gw150914: The advanced ligo detectors in the era of first
  discoveries.
\newblock {\em Phys. Rev. Lett.}, 116:131103, Mar 2016.

\bibitem{0264-9381-32-2-024001}
{The Virgo Collaboration} and F.~a et~al. {Acernese}.
\newblock Advanced virgo: a second-generation interferometric gravitational
  wave detector.
\newblock {\em Classical and Quantum Gravity}, 32(2):024001, 2015.

\bibitem{2018arXiv181112907T}
{The LIGO Scientific Collaboration}, {the Virgo Collaboration}, and B.~P.
  et~al. {Abbott}.
\newblock {GWTC-1: A Gravitational-Wave Transient Catalog of Compact Binary
  Mergers Observed by LIGO and Virgo during the First and Second Observing
  Runs}.
\newblock {\em arXiv e-prints}, page arXiv:1811.12907, November 2018.

\bibitem{Abbott20162}
{The LIGO Scientific Collaboration}, {the Virgo Collaboration}, and B.~P.
  et~al. {Abbott}.
\newblock {Observation of gravitational waves from a binary black hole merger}.
\newblock {\em Physical Review Letters}, 116(6):1--16, 2016.

\bibitem{2016PhRvL.116x1103A}
{The LIGO Scientific Collaboration}, {the Virgo Collaboration}, and B.~P.
  et~al. {Abbott}.
\newblock {GW151226: Observation of Gravitational Waves from a 22-Solar-Mass
  Binary Black Hole Coalescence}.
\newblock {\em Physical Review Letters}, 116(24):241103, jun 2016.

\bibitem{2017PhRvL.119n1101A}
{The LIGO Scientific Collaboration}, {the Virgo Collaboration}, and B.~P.
  et~al. {Abbott}.
\newblock {GW170814: A Three-Detector Observation of Gravitational Waves from a
  Binary Black Hole Coalescence}.
\newblock {\em Physical Review Letters}, 119(14):141101, oct 2017.

\bibitem{2017PhRvL.118v1101A}
{The LIGO Scientific Collaboration}, {the Virgo Collaboration}, and B.~P.
  et~al. {Abbott}.
\newblock {GW170104: Observation of a 50-Solar-Mass Binary Black Hole
  Coalescence at Redshift 0.2}.
\newblock {\em Physical Review Letters}, 118(22):221101, jun 2017.

\bibitem{TheLIGOScientificCollaboration2017}
{The LIGO Scientific Collaboration}, {the Virgo Collaboration}, and B.~P.
  et~al. {Abbott}.
\newblock {GW170608: Observation of a 19-solar-mass Binary Black Hole
  Coalescence}.
\newblock 35, 2017.

\bibitem{2017PhRvL.119p1101A}
{The LIGO Scientific Collaboration}, {the Virgo Collaboration}, and B.~P.
  et~al. {Abbott}.
\newblock {GW170817: Observation of Gravitational Waves from a Binary Neutron
  Star Inspiral}.
\newblock {\em Physical Review Letters}, 119(16):161101, oct 2017.

\bibitem{2004PhRvD..69h2004A}
{The LIGO Scientific Collaboration} and B.~P. et~al. {Abbott}.
\newblock {Setting upper limits on the strength of periodic gravitational waves
  from PSR J1939+2134 using the first science data from the GEO 600 and LIGO
  detectors}.
\newblock {\em \prd}, 69(8):082004, April 2004.

\bibitem{2005PhRvL..94r1103A}
B.~P. et~al. {The LIGO Scientific Collaboration}~{Abbott}.
\newblock {Limits on Gravitational-Wave Emission from Selected Pulsars Using
  LIGO Data}.
\newblock {\em Physical Review Letters}, 94(18):181103, May 2005.

\bibitem{2007PhRvD..76d2001A}
{The LIGO Scientific Collaboration}, {the Virgo Collaboration}, and B.~P.
  et~al. {Abbott}.
\newblock {Upper limits on gravitational wave emission from 78 radio pulsars}.
\newblock {\em \prd}, 76(4):042001, August 2007.

\bibitem{2008ApJ...683L..45A}
{The LIGO Scientific Collaboration}, {the Virgo Collaboration}, and B.~P.
  et~al. {Abbott}.
\newblock {Beating the Spin-Down Limit on Gravitational Wave Emission from the
  Crab Pulsar}.
\newblock {\em \apjl}, 683:L45, August 2008.

\bibitem{2010ApJ...713..671A}
{The LIGO Scientific Collaboration}, {the Virgo Collaboration}, and B.~P.
  et~al. {Abbott}.
\newblock {Searches for Gravitational Waves from Known Pulsars with Science Run
  5 LIGO Data}.
\newblock {\em \apj}, 713:671--685, April 2010.

\bibitem{2011ApJ...737...93A}
{The LIGO Scientific Collaboration}, {the Virgo Collaboration}, and J.~et~al.
  {Abadie}.
\newblock {Beating the Spin-down Limit on Gravitational Wave Emission from the
  Vela Pulsar}.
\newblock {\em \apj}, 737:93, August 2011.

\bibitem{2014ApJ...785..119A}
{The LIGO Scientific Collaboration}, {the Virgo Collaboration}, and J.~et~al.
  {Aasi}.
\newblock {Gravitational Waves from Known Pulsars: Results from the Initial
  Detector Era}.
\newblock {\em \apj}, 785:119, April 2014.

\bibitem{2017ApJ...839...12A}
{The LIGO Scientific Collaboration}, {the Virgo Collaboration}, and B.~P.
  et~al. {Abbott}.
\newblock {First Search for Gravitational Waves from Known Pulsars with
  Advanced LIGO}.
\newblock {\em \apj}, 839:12, April 2017.

\bibitem{Authors:2019ztc}
{The LIGO Scientific Collaboration}, {the Virgo Collaboration}, and B.~P.
  et~al. {Abbott}.
\newblock {Searches for Gravitational Waves from Known Pulsars at Two Harmonics
  in 2015-2017 LIGO Data}.
\newblock 2019.

\bibitem{2015ApJ...813...39A}
{The LIGO Scientific Collaboration}, {the Virgo Collaboration}, and J.~et~al.
  {Aasi}.
\newblock {Searches for Continuous Gravitational Waves from Nine Young
  Supernova Remnants}.
\newblock {\em \apj}, 813:39, November 2015.

\bibitem{2018arXiv181211656T}
{The LIGO Scientific Collaboration}, {the Virgo Collaboration}, and B.~P.
  et~al. {Abbott}.
\newblock {Searches for Continuous Gravitational Waves from Fifteen Supernova
  Remnants and Fomalhaut b with Advanced LIGO}.
\newblock {\em arXiv e-prints}, December 2018.

\bibitem{2015PhRvD..91f2008A}
{The LIGO Scientific Collaboration}, {the Virgo Collaboration}, and J.~et~al.
  {Aasi}.
\newblock {Directed search for gravitational waves from Scorpius X-1 with
  initial LIGO data}.
\newblock {\em \prd}, 91(6):062008, March 2015.

\bibitem{2017PhRvD..95l2003A}
{The LIGO Scientific Collaboration}, {the Virgo Collaboration}, B.~P. {Abbott},
  and {Abbott} et~al.
\newblock {Search for gravitational waves from Scorpius X-1 in the first
  Advanced LIGO observing run with a hidden Markov model}.
\newblock {\em \prd}, 95(12):122003, June 2017.

\bibitem{2017ApJ...847...47A}
{The LIGO Scientific Collaboration}, {the Virgo Collaboration}, and B.~P.
  et~al. {Abbott}.
\newblock {Upper Limits on Gravitational Waves from Scorpius X-1 from a
  Model-based Cross-correlation Search in Advanced LIGO Data}.
\newblock {\em \apj}, 847:47, September 2017.

\bibitem{PhysRevLett.120.031104}
{The LIGO Scientific Collaboration}, {the Virgo Collaboration}, and B.~P.
  et~al. {Abbott}.
\newblock First search for nontensorial gravitational waves from known pulsars.
\newblock {\em Phys. Rev. Lett.}, 120:031104, Jan 2018.

\bibitem{2015PhRvD..91b2004A}
{The LIGO Scientific Collaboration}, {the Virgo Collaboration}, and J.~et~al.
  {Aasi}.
\newblock {Narrow-band search of continuous gravitational-wave signals from
  Crab and Vela pulsars in Virgo VSR4 data}.
\newblock {\em \prd}, 91(2):022004, January 2015.

\bibitem{2017PhRvD..96l2006A}
{The LIGO Scientific Collaboration}, {the Virgo Collaboration}, and B.~P.
  et~al. {Abbott}.
\newblock {First narrow-band search for continuous gravitational waves from
  known pulsars in advanced detector data}.
\newblock {\em \prd}, 96(12):122006, December 2017.

\bibitem{doi:10.1046/j.1365-8711.2002.05180.x}
D.~I. Jones and N.~Andersson.
\newblock Gravitational waves from freely precessing neutron stars.
\newblock {\em Monthly Notices of the Royal Astronomical Society},
  331(1):203--220, 2002.

\bibitem{2010CQGra..27s4016A}
P.~{Astone}, S.~{D'Antonio}, S.~{Frasca}, and C.~{Palomba}.
\newblock {A method for detection of known sources of continuous gravitational
  wave signals in non-stationary data}.
\newblock {\em Classical and Quantum Gravity}, 27(19):194016, October 2010.

\bibitem{1998PhRvD..58f3001J}
P.~{Jaranowski}, A.~{Kr{\'o}lak}, and B.~F. {Schutz}.
\newblock {Data analysis of gravitational-wave signals from spinning neutron
  stars: The signal and its detection}.
\newblock {\em \prd}, 58(6):063001, September 1998.

\bibitem{2013A&A...552A..59B}
M.~{Bejger}.
\newblock {Parameters of rotating neutron stars with and without hyperons}.
\newblock {\em \aap}, 552:A59, April 2013.

\bibitem{2014PhRvD..89f2008A}
P.~{Astone}, A.~{Colla}, S.~{D'Antonio}, S.~{Frasca}, C.~{Palomba}, and
  R.~{Serafinelli}.
\newblock {Method for narrow-band search of continuous gravitational wave
  signals}.
\newblock {\em \prd}, 89(6):062008, March 2014.

\bibitem{2017CQGra..34m5007M}
S.~{Mastrogiovanni}, P.~{Astone}, S.~{D'Antonio}, S.~{Frasca}, G.~{Intini},
  P.~{Leaci}, A.~{Miller}, C.~{Palomba}, O.~J. {Piccinni}, and A.~{Singhal}.
\newblock {An improved algorithm for narrow-band searches of continuous
  gravitational waves}.
\newblock {\em Classical and Quantum Gravity}, 34(13):135007, July 2017.

\bibitem{2017PhRvD..96j2001C}
C.~{Cahillane}, J.~{Betzwieser}, D.~A. {Brown}, E.~{Goetz}, E.~D. {Hall},
  K.~{Izumi}, S.~{Kandhasamy}, S.~{Karki}, J.~S. {Kissel}, G.~{Mendell}, R.~L.
  {Savage}, D.~{Tuyenbayev}, A.~{Urban}, A.~{Viets}, M.~{Wade}, and A.~J.
  {Weinstein}.
\newblock {Calibration uncertainty for Advanced LIGO's first and second
  observing runs}.
\newblock {\em \prd}, 96(10):102001, November 2017.

\bibitem{2005AJ....129.1993M}
R.~N. {Manchester}, G.~B. {Hobbs}, A.~{Teoh}, and M.~{Hobbs}.
\newblock {The Australia Telescope National Facility Pulsar Catalogue}.
\newblock {\em \aj}, 129:1993--2006, April 2005.

\bibitem{2017ApJ...835...29Y}
J.~M. {Yao}, R.~N. {Manchester}, and N.~{Wang}.
\newblock {A New Electron-density Model for Estimation of Pulsar and FRB
  Distances}.
\newblock {\em \apj}, 835:29, January 2017.

\bibitem{2013AandA...560A..18K}
R.~{Kothes}.
\newblock {Distance and age of the pulsar wind nebula 3C 58}.
\newblock {\em \aap}, 560:A18, December 2013.

\bibitem{2008ApJ...677.1201K}
D.~L. {Kaplan}, S.~{Chatterjee}, B.~M. {Gaensler}, and J.~{Anderson}.
\newblock {A Precise Proper Motion for the Crab Pulsar, and the Difficulty of
  Testing Spin-Kick Alignment for Young Neutron Stars}.
\newblock {\em \apj}, 677:1201--1215, April 2008.

\bibitem{2013Natur.495...76P}
G.~{Pietrzy{\'n}ski}, D.~{Graczyk}, W.~{Gieren}, I.~B. {Thompson},
  B.~{Pilecki}, A.~{Udalski}, I.~{Soszy{\'n}ski}, S.~{Koz{\l}owski},
  P.~{Konorski}, K.~{Suchomska}, G.~{Bono}, P.~G.~P. {Moroni}, S.~{Villanova},
  N.~{Nardetto}, F.~{Bresolin}, R.~P. {Kudritzki}, J.~{Storm}, A.~{Gallenne},
  R.~{Smolec}, D.~{Minniti}, M.~{Kubiak}, M.~K. {Szyma{\'n}ski}, R.~{Poleski},
  {\L}.~{Wyrzykowski}, K.~{Ulaczyk}, P.~{Pietrukowicz}, M.~{G{\'o}rski}, and
  P.~{Karczmarek}.
\newblock {An eclipsing-binary distance to the Large Magellanic Cloud accurate
  to two per cent}.
\newblock {\em \nat}, 495:76--79, March 2013.

\bibitem{2003ApJ...596.1137D}
R.~{Dodson}, D.~{Legge}, J.~E. {Reynolds}, and P.~M. {McCulloch}.
\newblock {The Vela Pulsar's Proper Motion and Parallax Derived from VLBI
  Observations}.
\newblock {\em \apj}, 596:1137--1141, October 2003.

\bibitem{2012ApJ...755...39V}
J.~P.~W. {Verbiest}, J.~M. {Weisberg}, A.~A. {Chael}, K.~J. {Lee}, and D.~R.
  {Lorimer}.
\newblock {On Pulsar Distance Measurements and Their Uncertainties}.
\newblock {\em \apj}, 755:39, August 2012.

\bibitem{2010ApJ...716..663R}
M.~{Renaud}, V.~{Marandon}, E.~V. {Gotthelf}, J.~{Rodriguez}, R.~{Terrier},
  F.~{Mattana}, F.~{Lebrun}, J.~A. {Tomsick}, and R.~N. {Manchester}.
\newblock {Discovery of a Highly Energetic Pulsar Associated with IGR
  J14003-6326 in the Young Uncataloged Galactic Supernova Remnant G310.6-1.6}.
\newblock {\em \apj}, 716:663--670, June 2010.

\bibitem{2014ApJ...795..168M}
M.~{Marelli}, A.~{Harding}, D.~{Pizzocaro}, A.~{De Luca}, K.~S. {Wood},
  P.~{Caraveo}, D.~{Salvetti}, P.~M. {Saz Parkinson}, and F.~{Acero}.
\newblock {On the Puzzling High-Energy Pulsations of the Energetic Radio-Quiet
  {$\gamma$}-Ray Pulsar J1813-1246}.
\newblock {\em \apj}, 795:168, November 2014.

\bibitem{2012ApJ...753L..14H}
J.~P. {Halpern}, E.~V. {Gotthelf}, and F.~{Camilo}.
\newblock {Spin-down Measurement of PSR J1813-1749: The Energetic Pulsar
  Powering HESS J1813-178}.
\newblock {\em \apjl}, 753:L14, July 2012.

\bibitem{2006ApJ...637..456C}
F.~{Camilo}, S.~M. {Ransom}, B.~M. {Gaensler}, P.~O. {Slane}, D.~R. {Lorimer},
  J.~{Reynolds}, R.~N. {Manchester}, and S.~S. {Murray}.
\newblock {PSR J1833-1034: Discovery of the Central Young Pulsar in the
  Supernova Remnant G21.5-0.9}.
\newblock {\em \apj}, 637:456--465, January 2006.

\bibitem{2008ApJ...681..515G}
E.~V. {Gotthelf} and J.~P. {Halpern}.
\newblock {Discovery of a Young, Energetic 70.5 ms Pulsar Associated with the
  TeV Gamma-Ray Source HESS J1837-069}.
\newblock {\em \apj}, 681:515--521, July 2008.

\bibitem{2014ApJ...790..103A}
P.~{Arumugasamy}, G.~G. {Pavlov}, and O.~{Kargaltsev}.
\newblock {XMM-Newton Observations of Young and Energetic Pulsar J2022+3842}.
\newblock {\em \apj}, 790:103, August 2014.

\bibitem{2016MNRAS.458.3341D}
G.~{Desvignes}, R.~N. {Caballero}, L.~{Lentati}, J.~P.~W. {Verbiest}, D.~J.
  {Champion}, B.~W. {Stappers}, G.~H. {Janssen}, P.~{Lazarus},
  S.~{Os{\l}owski}, S.~{Babak}, C.~G. {Bassa}, P.~{Brem}, M.~{Burgay},
  I.~{Cognard}, J.~R. {Gair}, E.~{Graikou}, L.~{Guillemot}, J.~W.~T. {Hessels},
  A.~{Jessner}, C.~{Jordan}, R.~{Karuppusamy}, M.~{Kramer}, A.~{Lassus},
  K.~{Lazaridis}, K.~J. {Lee}, K.~{Liu}, A.~G. {Lyne}, J.~{McKee}, C.~M.~F.
  {Mingarelli}, D.~{Perrodin}, A.~{Petiteau}, A.~{Possenti}, M.~B. {Purver},
  P.~A. {Rosado}, S.~{Sanidas}, A.~{Sesana}, G.~{Shaifullah}, R.~{Smits}, S.~R.
  {Taylor}, G.~{Theureau}, C.~{Tiburzi}, R.~{van Haasteren}, and A.~{Vecchio}.
\newblock {High-precision timing of 42 millisecond pulsars with the European
  Pulsar Timing Array}.
\newblock {\em \mnras}, 458:3341--3380, May 2016.

\bibitem{2001ApJ...552L.125H}
J.~P. {Halpern}, F.~{Camilo}, E.~V. {Gotthelf}, D.~J. {Helfand}, M.~{Kramer},
  A.~G. {Lyne}, K.~M. {Leighly}, and M.~{Eracleous}.
\newblock {PSR J2229+6114: Discovery of an Energetic Young Pulsar in the Error
  Box of the EGRET Source 3EG J2227+6122}.
\newblock {\em \apjl}, 552:L125--L128, May 2001.

\bibitem{2015PhRvD..91j2003L}
P.~{Leaci} and R.~{Prix}.
\newblock {Directed searches for continuous gravitational waves from binary
  systems: Parameter-space metrics and optimal Scorpius X-1 sensitivity}.
\newblock {\em \prd}, 91(10):102003, May 2015.

\bibitem{2018Natur.556..219P}
J.~{Palfreyman}, J.~M. {Dickey}, A.~{Hotan}, S.~{Ellingsen}, and W.~{van
  Straten}.
\newblock {Alteration of the magnetosphere of the Vela pulsar during a glitch}.
\newblock {\em \nat}, 556:219--222, April 2018.

\bibitem{2018PhRvD..97h2002C}
P.~B. {Covas}, A.~{Effler}, E.~{Goetz}, P.~M. {Meyers}, A.~{Neunzert},
  M.~{Oliver}, B.~L. {Pearlstone}, V.~J. {Roma}, R.~M.~S. {Schofield}, V.~B.
  {Adya}, and et~al.
\newblock {Identification and mitigation of narrow spectral artifacts that
  degrade searches for persistent gravitational waves in the first two
  observing runs of Advanced LIGO}.
\newblock {\em \prd}, 97(8):082002, April 2018.

\bibitem{2013PhRvD..88d4016J}
Nathan~K. {Johnson-McDaniel}.
\newblock {Gravitational wave constraints on the shape of neutron stars}.
\newblock {\em \prd}, 88(4):044016, Aug 2013.

\bibitem{2013A&A...560A..18K}
R.~{Kothes}.
\newblock {Distance and age of the pulsar wind nebula 3C 58}.
\newblock {\em \aap}, 560:A18, December 2013.

\end{thebibliography}

\end{document}